\input amstex
\documentstyle{amsppt}
\magnification =\magstep1
\hsize 6.5truein
\vsize 9truein
\addto\tenpoint{\normalbaselineskip=20pt\normalbaselines}
\expandafter\redefine\csname logo\string@\endcsname{}
\topmatter
\title General Balance Functions \\
in the Theory of Interest
\endtitle
\rightheadtext{General Balance Functions in the Theory of
Interest}
\author David Spring\endauthor
\affil Department of Mathematics, Glendon College, York
University, Canada\endaffil \abstract We develop an axiomatic
theory of balance functions (future value functions) in the theory
of interest that is derived from financial considerations and
which applies to general regulated payment streams, including
continuous payment streams. Balance functions exist and are unique
up to an initial choice of deposit and investment accumulation
functions. In terms of these balance functions we also construct a
unique internal rate of return for each regulated payment stream
that is an investment project. This theory subsumes and clarifies
previous theories of internal rate of return functions for more
specialized classes of investment projects.

\vskip .25cm \flushpar{\smc Key Words:} balance function;
investment project; internal rate of return \vskip .25cm \flushpar
\endabstract
\endtopmatter
\TagsOnLeft
\define\les{\le}
\define\ges{\ge}

\define\RK{\Cal R_K}
\define\SK{\Cal S_K}
\define\IK{\Cal I_K}
\define\jc{\Cal B_j(f_{\Cal C})}
\define\jd{\Cal B_j(f_{\Cal D})}
\define\bc1{\Cal B_{j +1}(f)}
\define\bd1{B_{j +1}(\Cal D)}

\define\ep{\epsilon}
\define\dj{\Delta_j}
\define\cd{\Delta_{j +1}}
\define\gy{\Cal B_j(f_{\Cal C -\Cal D})(y,0)}
\define\ly{\Cal B_j(f_{\Cal C -\Cal D})(0,y)}
\define\at{a(t_j,t_{j +1})}
\define\bt{b(t_j,t_{j +1})}
\define\yt{y(t_j,t_{j +1})}

\define\rt{\: \bold R\times\Cal R\to\bold R}
\count 0=1

\document
\footnote""{\flushpar {\it Address}\,: Professor David Spring,
Department of Mathematics, Glendon College, York University, 2275 Bayview Avenue,
Toronto, Ontario, Canada, M4N 3M6.} \footnote""{\flushpar {\it
Tel}\,: 416 487 6815. {\it Fax}\,: 416 487 6852.}
\footnote""{\flushpar {\it email}\,: dspring\@glendon.yorku.ca}

\newpage

\centerline{\bf 1. Introduction} \vskip .25cm

\flushpar{\bf 1.1.} A historical problem in mathematical finance
is the determination of an internal rate of return (IRR) for a
general investment project. The financial motivation is that
portfolios with a high IRR make for more attractive investment
choices. The special case of determining the IRR of a loan
contract, i.e., an initial cash outflow followed by a finite
sequence of cash inflows, is a well-known computation in the
financial mathematics literature. Briefly, the present value
function is a strictly decreasing function of the rate per period
$i$ whose unique root is the IRR of the loan contract, cf. Donald
(1970), Kellison (1991). Indeed, the computation of the IRR in the
case of loan contracts (mortgages, bonds, etc.) constitutes one of
the main applications historically of financial mathematics to
problems in the business world. However general investment
projects are not loan contracts since they typically involve
positive cash flows (inflows) interspersed with negative cash
flows (outflows); it is the occurrence of large swings in the
sequence of inflows and outflows that is responsible for the
failure of the present value method to determine a unique IRR for
a general investment project: in general multiple roots must
occur. This rather awkward situation has led over the years to a
search for more general methods for computing the IRR of
investment projects that include as a special case the classical
IRR computation for loan contracts. Notable among these
generalizations are the following:

\flushpar{\bf(i)} Arrow and Levhari (1969) define a unique IRR
$r_f$ for an investment project $f$ that is defined by a
continuous payment stream with finite time horizon (duration) that
also is differentiable and changes sign only a finite number of
times. For a constant rate of discount, these authors consider the
maximum of the present values of the project $f$ calculated over
all truncated time periods, i.e., over all initial time intervals
of the project $f$. The authors' key observation is that this
maximized present value is a monotone decreasing function of the
rate of discount. The unique root of this decreasing function is
defined to be the IRR $r_f$ of the investment project $f$. The IRR
$r_f$ coincides with the classical IRR in case the investment
project $f$ is a discrete loan contract.

The decreasing property of the above maximized present value
function is a consequence of the well-known argument of ``positive
tails,'' perhaps first employed in the economic literature by
Wright (1959) in the case of finite cash flow sequences; further
details are provided in Promislow and Spring (1996, Appendix).
Refinements of the approach of Arrow and Levhari (1969) are
developed in Fleming and Wright (1971), Sen (1975).

\flushpar{\bf (ii)} A completely different solution to the IRR
problem, of importance to this paper, was proposed by Teichroew et
al.,(1965a,b) in the context \underbar{only} of investment
projects $f$ defined by finite cash flow sequences. These authors
begin with the financial observation that their are \underbar{two}
types of interest rates: a {\it deposit} rate of interest that
applies to current balances that are positive, i.e., current
surpluses; an {\it investment} rate of interest that applies to
current balances that are negative, i.e., current debts. In
practice the current deposit rate is less than the current
investment rate. According to these authors the current balance
$\Cal B_j(f)$ of an investment project $f$, at the time $t_j$ of
the $j$th cash flow $C_j$ of the project $f$, should be calculated
inductively by applying the deposit interest rate, respectively
the investment interest rate, to the previous current balance
$\Cal B_{j-1}(f)$ during the $j$th period $[t_{j-1},t_j]$,
according to whether the balance $\Cal B_{j-1}(f)\ge 0$ (a
surplus), respectively $B_{j-1}(f)\le 0$ (a debt). In the special
case that the deposit and investment rates of interest coincide
then the successive balances $\Cal B_j(f)$ are just the classical
accumulated value functions at times $t_j$ determined by this
common interest rate. For a fixed deposit rate $i$ per period
during the life of the project, these authors crucially observe
that each current balance $\Cal B_j(f)$, calculated inductively as
above, is a monotone decreasing function of the investment rate of
interest. The IRR $i_f$ of an investment project $f$ is then
defined to be the unique root of the monotone decreasing balance
function $\Cal B_n(f)$ calculated at end of the investment project
(at the time $t_n$ of the last cash flow $C_n$ of the project).
The IRR $i_f$ also coincides with the classical IRR in case the
investment project $f$ is a discrete loan contract.

This ``two-interest-rate'' theory for calculating current balances
circumvents the problem of multiple roots that occur in the
present value method for determining the IRR. Multiple roots occur
because of the implicit assumption, false for economic reasons,
that the above deposit and investment rates of interest coincide.
Standard texts, Kellison (1991), discuss with examples the IRR due
to Teichroew et al.,(1965a,b).

Despite the different financial points of view outlined briefly in
(i), (ii) above, Promislow and Spring (1996) prove somewhat
surprisingly that the IRR $r_f$ due to Arrow and Levhari (1969),
in the context of finite cash flow sequences, is a special case of
the IRR $i_f$ due to Teichroew et al.,(1965a,b) when the deposit
rate per period $i$ tends to infinity: $r_f=\lim_{i\to\infty}i_f$.
In this sense the IRR $i_f$ is more general that the IRR $r_f$,
and depends on an analysis of balance functions in terms of
deposit and investment rates of interest.

\flushpar{\bf 1.2.} An open question in the literature, related to
the above issues, is whether the IRR of Teichroew et al.,\,(1965a)
can be extended to the general case of investment projects defined
by continuous payment streams, and if so whether the IRR of Arrow
and Levhari (1969) will again be a limiting case in this general
context. This generalization requires a suitable theory of balance
functions for continuous payment streams. A major obstacle to this
generalization is that the theory of Teichroew et al.,\,(1965a)
treats only finite cash flow sequences for which balance functions
are defined inductively and depend importantly on the sign of the
previously defined balance function. This inductive procedure to
define balance functions cannot apply to continuous payment
streams. In addition the work of Promislow and Spring (1996)
suggests that the IRR of Teichroew et al.,\,(1965a) should be
universal with respect to IRR functions defined in terms of
balance functions. These questions form the subject matter of this
paper. We remark here that the question of an expectation value
for the IRR, in the stochastic setting, has yet to be addressed in
this burgeoning new area of financial mathematics. This
presupposes a clearer understanding in the financial literature of
the determination of the IRR in the classical setting of general
investment projects subject to deterministic accumulation
functions. The results of our paper contribute towards a better
understanding of this classical situation.

\flushpar{\bf 1.3.} In this paper we solve the balance function
problem posed above in \S1.2 by developing a new theory of balance
functions (\S4), expressed in axiomatic terms, that is
sufficiently general to treat payment streams that are regulated
functions of ``finite time horizon'' i.e., regulated payment
streams supported on compact time intervals (\S3.2). Continuous
payment streams and step function payment streams are important
special cases. Mathematically, a function $f$ on a compact
interval $[a,b]$ is {\it regulated} if $f=\lim_{n\to\infty}f_n$
where each $f_n$ is a step function on $[a,b]$ (step function
payment streams correspond to finite cash flow sequences), and
where the limit is taken in the topology of uniform convergence of
functions on the compact interval $[a,b]$.  Let $\Cal
B_t(f)\in\bold R$ denote the balance of the payment stream $f$ at
time $t$. The continuity axiom $\Cal A_5$ states that the balance
$\Cal B_t(f)=\lim_{n\to\infty}\Cal B_t(f_n)$, where
$f=\lim_{n\to\infty}f_n$ as above. In this way balance functions
on the space of step function payment streams extend by the
continuity axiom to balance functions on the space of regulated
payment streams. This is the essence of our topological approach
to the theory of balance functions on general regulated payment
streams. Implicit in this topological approach is the development
of analytic estimates that ensure the convergence properties of
the limit in axiom $\Cal A_5$. Some of these analytic estimates
are rather lengthy, as in the proof of Theorem 5.3, and therefore
are relegated to the Appendix.

We note here that the axioms for balance functions (\S4) allow one
to reconstruct deposit and investment accumulation functions in
the spirit of Teichroew et al.,(1965a). Briefly, let $a(s,t)\ge 0$
be the balance at time $t$ of a single cash flow of 1 (deposit of
1 unit) at time $s$, $s\le t$; similarly let $b(s,t)\ge 0$ be the
negative of the balance at time $t$ of a single cash flow of $-1$
(debt of 1 unit) at time $s$, $s\le t$. It follows from Lemma 4.1
that $a(s,t)$, $b(s,t)$ are accumulation functions (\S2), denoted
as deposit, respectively, investment accumulated functions. In
general $a(s,t)$, $b(s,t)$ are distinct accumulation functions, in
conformity with theory developed by Teichroew et al., (1965a,b).
Indeed, a special case is $a(s,t)=(1+\alpha)^{t-s}$,
$b(s,t)=(1+\beta)^{t-s}$, where $\alpha$, respectively $\beta$, is
the deposit rate, respectively the investment rate, per period
that was introduced in Teichroew et al.,(1965a). In this way the
axioms code for a ``two-interest-rate'' general theory of balance
functions that applies for example to continuous payment streams.

Theorem 4.5 proves that balance functions that satisfy the axioms
do exist and are unique up to initial choices of deposit and
investment accumulation functions. Furthermore, in terms of these
balance functions, there is a natural way to define the IRR of all
investment projects, including as a special case the IRR of
Teichroew et al.,\,(1965a); indeed, our construction of IRR
functions is inspired by this special case, thus solving the IRR
problem posed in \S1.2. In this sense the IRR of Teichroew et
al.,\,(1965a) is seen to have a universal character since it
occurs naturally in the context of balance functions that
themselves are uniquely determined axiomatically by {\it a priori}
financial considerations.

In somewhat more detail, let $a(s,t)$ be a fixed positive
(deposit) accumulation function of bounded variation (\S2.2); in
our theory $a(s,t)$ applies to current balances that are $\ges 0$
(surpluses) at time $s$. Let $x^{t-s}$, $x\ges 0$, be a variable
(investment) accumulation function; in our theory $x^{t-s}$
applies to current balances that are $\les 0$ (debt) at time $s$.
The main result Theorem 5.3 proves that for each investment
project $f$ (\S3.3) the balance function $\Cal B^{\,x}_d(f)$,
calculated at the time $d$ at the end of the investment project
$f$, (the dependence on $a(s,t)$ is omitted), is a strictly
decreasing function of $x$ such that $\lim_{x\to\infty}\Cal
B^{\,x}_d(f)=-\infty$, and therefore has at most one root
$x=1+i_f\ges 0$. The IRR of $f$ is defined to be the parameter
$i_f\ges -1$. If there is no root then $i_f=-1$. (cf. \S5.2 for
precise details). Although not shown here, if for example
$a(s,t)=(1+i)^{t-s}$, $i\ges -1$, then again
$\lim_{i\to\infty}(i_f) =r_f$, the IRR defined by Arrow and
Levhari (1969), in case $f$ is also a continuous payment stream.

The IRR $i_f$ coincides with the IRR of Teichroew et al.,\,(1965)
in case $f$ corresponds to a finite number of cash flows (the unit
of time is 1 period). In particular, $i_f$ equals the classical
IRR in case $f$ corresponds to a loan contract (mortgages, bonds
etc.). To summarize, we propose in this paper an axiomatic theory
of balance functions, in terms of which we define an IRR that
provides a comprehensive solution to the historical problem of
defining an IRR for general investment projects which occur in
mathematical economics and finance.

The generality of our approach requires the development {\it ab
initio} of the theory of accumulation functions (\S2) and of
payment streams (\S3). While economic arguments are indicated
where appropriate, the estimates developed in \S4, \S5 to justify
our topological approach for proving the main results are
presented there in detail since there is no convenient reference
to the economic and mathematical literature for these types of
calculations.

 \vskip .25cm \centerline
{\bf 2. Accumulation Functions} \vskip .25cm \flushpar{\bf 2.1.}
Accumulation functions are basic to the theory of interest since
they relate, in mathematical terms, the value of invested capital
at any one date to its value at any subsequent date. In this
section we develop the theory of accumulation functions in a more
general setting than appears in the economic and financial
literature.

Let $\Bbb H =\{(s,t) \in \bold R^2 \mid s\les t\}$, the half-space
above the line $y=x$ in $\bold R^2$. A non-negative function $a\:
\Bbb H\to [0,\infty )$, denoted $a\ges 0$, is an {\it accumulation
function}  if $a(t,t) =1$ for all $t\in \bold R$, and if the
following multiplicative property is satisfied:
$$
a(r,s)\cdot a(s,t) =a(r,t) \quad \text{for all }r\les s\les
t.\tag2.1
$$
Accumulation functions are not assumed to be continuous, and the
value $a(s,t)=0$ is allowed. An extreme example is the zero
accumulation function: $0(t,t)=1$ and $0(s,t)=0$ for all $s<t$. In
financial terms $a(s,t)$ is the accumulated (future) value at time
$t$ of one monetary unit invested at time $s$, for all $s\le t$
(throughout, unless specified to the contrary, the conventional
time unit is 1 year). If the accumulation function $a(s,t)$ is
positive i.e., $a\: \Bbb H\to (0,\infty)$, denoted $a> 0$, then
$a(s,t)$ extends naturally to all of $\bold R^2$ (same notation)
by requiring (2.1) to hold universally:
$$
a(r,s)\cdot a(s,t) =a(r,t) \quad \text{for all }r,s,t. \tag2.2
$$
In particular, $a(s,t)\cdot a(t,s) = a(s,s) =1$; hence $a(t,s)
=1/a(s,t)$ for all $s,t$. Let $f\: \bold R \to (0,\infty)$ be the
positive function $f(t) =a(x_0 ,t)$ ($x_0$ is an arbitrary
reference point). Setting $r =x_0$ in (2), it follows that,
$$
a(s,t) =f(t)/f(s) \quad \text{for all }s,t. \tag2.3
$$
Furthermore, writing $f(t) =e^{\,g(t)}$, we obtain the standard
representation of positive accumulation functions,
$$
a(s,t) = e^{\,g(t) -g(s)} \quad \text{for all }s,t. \tag2.4
$$
Note that $g\:\bold R\to\bold R$ is unique up to addition of a
constant. In the classical theory of interest
$$
g(t)=\int_0^t\delta(u)\,du,
$$
where the continuous function $\delta(t)$ is called the force of
interest; in measure theoretic terms $\delta(t)$ is the density
function associated to $g(t)$. The classical example is $g(t)=rt$,
for which $a(s,t)=e^{r(t-s)}$, where $r$ is the rate of
continuously compounded interest.

If $a(s,t) > 0$ then $a(t,s) = 1/a(s,t)$ is the present value
(price) at time $s$ of one monetary unit at time $t\ges s$. If
$a(s,t)=0$ for $s<t$, then one monetary unit at time $s$ becomes
worthless (value 0) at time $t$. The multiplicative property (2.1)
ensures coherence of monetary values at all intermediate times. In
practical examples $a(s,t)=(1+i)^{t-s}$, where $i> -1$ is a
constant rate per period (the time unit is 1 period). The value at
time $t$ of one monetary unit at time $s$ is $(1+i)^{t-s}$, a
basic computation in financial mathematics.

A convenient equivalent formulation of accumulation functions is
in terms of real-valued functions defined on the set of all
compact intervals in $\bold R$: if $J =[s,t]$, $s\les t$, then
$a(J) =a(s,t)$. The multiplicative property (2.1) is then
expressed as follows.
$$
a(J\cup K) =a(J)\cdot a(K),\tag2.5
$$
where $J =[r,s]$, $K =[s,t]$ are adjacent compact intervals.
Accumulation functions are partially ordered in the obvious way:
$a\les b$ if and only if $a(J) \les b(J)$ for all compact
intervals $J =[s,t]$, $s \les t$. Evidently, the product
$a(J)\cdot b(J)$ of accumulation functions is an accumulation
function. \vskip.25cm

\flushpar{\bf 2.2. Monotone Accumulation Functions.} An
accumulation function $a(J)$ is {\it monotone increasing
(decreasing)} if $a(J)\les a(K)$ ($a(J)\ges a(K)$) for all nested
compact intervals $J\subseteq K$. Since $a(t,t) =1$ for all $t\in
\bold R$, it follows that if $a(J)$ is monotone increasing,
respectively decreasing, then $a \ges 1$, hence positive,
respectively $a\le 1$. Evidently, if $y_1(J), y_2(J)$ are monotone
increasing, respectively decreasing, accumulation functions then
the product accumulation function $y(J)=y_1(J)\cdot y_2(J)$ is
also monotone increasing, and $y\ges y_1, y\ges y_2$ (a common
upper bound), respectively $y\les y_1, y\les y_2$ (a monotone
decreasing common lower bound). A positive accumulation function
$a(s,t) =\exp (g(t) -g(s))$ is monotone increasing if and only if
the function $g\: \bold R \to \bold R$ is monotone increasing.

An accumulation function is positive on an interval $I$ if $a(s,t)
>0$ for all $s,t\in I$, written $a> 0$ on $I$. As in (2.4), $a(s,t)
=e^{g(t) -g(s)}$ for all $s,t \in I$, where $g\: I\to \bold R$.
The positive accumulation function $a(s,t)$ is defined to be of
{\it bounded variation} on the compact interval $I$ if the
corresponding function $g\: I\to \bold R$ is of bounded variation.
For example, monotone increasing accumulation functions are of
bounded variation.

Let $a(s,t)=e^{\,g(t)-g(s)}$ be an accumulation function of
bounded variation on a compact interval $I$. Let $V_g([s,t])$ be
the variation function associated to $g$, defined on all intervals
$[s,t] \subseteq I$. Thus $V_g$ is a finitely additive interval
function and is monotone increasing: $V_g(J) \les V_g(K)$ for all
subintervals $J\subseteq K\subseteq I$. Note that $|g(t)-g(s)|\les
V_g(s,t)$ for all subintervals $[s,t]\subseteq I$. Let $h\: I\to
\bold R$ be the monotone increasing function $h(t)=V(c,t)$, $t\in
I$. Then $y(s,t)=e^{h(t)-h(s)}=e^{V_g(s,t)}$ is a monotone
increasing accumulation function on $I$ such that $a\les y$ on
$I$. Similarly, $x(s,t)=e^{h(s)-h(t)}=e^{-V(s,t)}$ is monotone
decreasing and $a\ges x$ on $I$. Conversely one can prove that if
$a>0$ and $a\les y$ where $y$ is monotone increasing, then $a$ is
of bounded variation on all compact intervals.\vskip.25cm

\centerline{\bf 3. Payment Streams} \vskip .25cm \flushpar{\bf
3.1. Regulated functions.} As discussed in the introduction, the
classical theory of interest has no framework for defining balance
functions of type Teichroew et al.,(1965) in the case of
continuous payment streams. Our general theory of balance
functions applies most naturally to payment streams that are
regulated functions. These include all payment streams of
theoretical and of practical interest, such as continuous payment
streams and step function payment streams associated to finite
cash flow sequences.

A function $f\: \bold R \to \bold R$ is {\it regulated} if $f$ has
finite right-hand and left-hand limits at each $t \in \bold R$. It
is well-known, Bourbaki (1949, Ch.II,\S1.3), Dieudonn\'e (1960),
that a function $f\: \bold R \to \bold R$ is regulated if and only
if on each compact interval $[a,b] \subset \bold R$, $f$ is the
limit of step functions in the topology of uniform convergence on
$[a,b]$. Employing pointwise right- and left-hand limits, it is
clear that if $f,g$ are regulated then the functions $f +g$,
$f\cdot g$ are regulated. The set of regulated functions on a
compact interval strictly includes the sets of step functions,
continuous functions, monotone functions and hence also the set of
functions of bounded variation. (A function of bounded variation
can be expressed as a difference of monotone functions.)

Regulated functions occur in the classical theory of interest in
the special case of step function payment streams associated to
finite cash flow sequences of the form $\Cal C =(C_i)_{0\les i\les
n}$ such that the cash flow $C_i$ occurs at time $t_i \in \bold
R$, $t_0 < t_1 \dots < t_n$. Associated to $\Cal C$ is the step
function $f_{\Cal C} \: \bold R \to \bold R$, continuous on the
right,
$$
f_{\Cal C} (t) =\sum_{t_i \les t} C_i. \tag3.1
$$
Hence $f_{\Cal C} (t) =0$ for all $t < t_0$ and $f_{\Cal C} (t)$
is constant $=\sum_{i} C_i$ for all $t\ges t_n$. Thus $C_0 =f_{\Cal
C} (t_0)$, and if $i\ges 1$, the cash flow $C_i$ is the difference,
$$
C_i =f_{\Cal C} (t_i) -f_{\Cal C} (t_{i -1}), \quad 1\les i\les n.
\tag3.2
$$
The cash flow $C_i$ represents the jump of the step function
$f_{\Cal C}$ at time $t_i$, $0\les i\les n$. Conversely, let $f\:
\bold R \to \bold R$ be a step function, continuous on the right,
such that $f(t) =0$ on some interval $(-\infty ,a)$. Evidently
there is a finite cash flow sequence $\Cal C =(C_i)_{0\les i\les n}$
such that $f=f_{\Cal C}$. Continuity on the right is the standard
convention in financial mathematics which implies that a cash flow
payment is at the receivers disposal immediately as it falls due
and thereafter.

If $s < t$ note that $f_{\Cal C} (t) - f_{\Cal C} (s)$ is the sum
of the cash flows in the interval $(s,t]$; hence the terminology
that $f_{\Cal C}$ is a {\it distribution function}\,: for each
$t$, $f_{\Cal C}(t)$ is the sum of all the cash flows on the
interval $(-\infty, t]$.

In this paper we develop the theory of interest based on regulated
payment stream functions $f\: \bold R \to \bold R$ which have
compact support, \S3.2. Our strategy is to prove general theorems
in the case of step function payment streams of the type (3.1)
above. An important feature is our topological approach: The main
constructs (balance functions, internal rates of return etc.) are
defined first on the space of step functions. Since step functions
are dense in the space of compactly supported regulated functions
(in the topology of uniform convergence), the corresponding
constructs in the case of regulated payment stream functions are
defined topologically by passing to the uniform limit.

\flushpar{\bf 3.2. Regulated Payment Streams}

\flushpar A regulated {\it payment stream}, or flow function, is a
regulated function $f\: \bold R \to \bold R$, continuous on the
right, which is supported in a compact interval in the following
sense: there is a compact interval $[a,b] \subset \bold R$, $a\les
b$ (which depends on $f$) such that $f = 0$ on the interval
$(-\infty ,a)$ and $f$ is constant on the interval $[b, \infty)$.
The intersection over all compact intervals on which $f$ is
supported is the {\it minimal support} of the regulated payment
stream $f$. Note that the minimal support is empty only in the
extreme case that $f =0$ on $\bold R$. The canonical example of a
regulated payment stream is the step function (3.1), $f_{\Cal C}
\: \bold R \to \bold R$, associated to a finite cash flow sequence
$\Cal C =(C_i)_{0\les i\les n}$, supported in the interval $[t_0
,t_n]$; this interval is the minimal support if and only if $C_0 ,
C_n$ are both non-zero. As explained above, continuity on the
right is the conventional requirement for payment streams in
financial mathematics. Let $\Cal R$ be the set of all regulated
payment streams $f\: \bold R\to\bold R$. For each compact interval
$K\subset \bold R$, let $\RK \subset \Cal R$ be the subset of
regulated payment streams whose minimal support is contained in
$K$. If $K\subseteq L$, then $\RK \subseteq \Cal R_L$, and $\Cal
R=\bigcup_K \RK$. Let $\Cal S\subset\Cal R$ be the subset of step
functions, and define $\SK=\Cal S\cap \RK$.

For each $K$, $\SK$ is dense in $\RK$ in the topology of uniform
convergence. To see this, let $g\in \RK$ and let $\ep
> 0$. Since $g$ has left and right hand limits at each point we
observe that for each $t$ there is function $h\:
I=(t-\eta,t+\eta)\to \bold R$ such that: $h(t)=g(t)$; $h$ is
constant on each interval $(t-\eta,t)$, $[t,t+\eta)$ (in
particular $h$ is right continuous); $|h-g|_I<\ep$. Since $K$ is
compact, employing the observation, there is a partition of the
interval $K =[u,v]$, $t_0=u < t_1< \dots < t_n=v$, and a step
function $f \in \SK$ such that: \item{(i)} $f(t_i) =g(t_i)$,
$0\les i\les n$; $f(t)=0$, $t< u$; $f(t)=g(v)$, $t\ges v$.
\item{(ii)} $f(t)$ is constant $=g(t_i)$ on the interval $[t_i
,t_{i +1})$, $1\les i \les n -1$. \item{(iii)} $|f(t) -g(t)| <\ep$
for all $t\in\bold R$.

Employing (3.1), (3.2), $f =f_{\Cal C}\in\Cal S_K$, where $\Cal C
=(C_i)_{0\les i\les n}$, is defined as follows:
$$
C_0 =g(t_0)\,;\quad C_i =g(t_i) -g(t_{i -1}) \quad 1\les i\les
n.\tag3.3
$$
Property (iii) shows that $\SK$ is dense in $\RK$. Employing (3.3)
we remark also that all the cash flows of $\Cal C$, for the
approximating step function $f_{\Cal C}$, are $\les 0$,
respectively $\ges 0$, if $g(t_0)\les 0$ and $g$ is monotone
decreasing, respectively $g(t_0)\ges 0$ and $g$ is monotone
increasing.

Let $f\in\RK$ and let $(f_n)_{n\ges 1}$ be a Cauchy sequence in
$\SK$ such that $f=\lim_{n\to\infty}f_n$ uniformly on $K$. For
each $n$ the distribution function $f_n(t)$ is the sum of all the
corresponding cash flows in the interval $(-\infty,t]$.
Consequently, in the uniform limit, the payment stream $f(t)$ also
is viewed as a distribution function which for each $t\in\bold R$
is the ``total cash flow'' in the interval $(-\infty,t]$. To
explain this, suppose in addition $f\in\RK$ is of bounded
variation on $K$, hence of bounded variation on each compact
interval in $\bold R$. For each $t\in\bold R$,
$$
f(t)=\int^t_{-\infty}df,
$$
where the Stieltjes integral is employed (cf. Promislow (1980)).
Furthermore if $f\in\RK$ is a continuous payment stream of class
$C^1$ then,
$$
f(t)=\int_{-\infty}^tdf=\int^t_{-\infty}f^\prime dt.\tag3.4
$$
In this context $f^\prime(t)$ is the signed density function of
the function $f(t)$. Following the common practice in applied
mathematics for interpreting Riemann integrals, it is still
current in the financial and economics literature, Arrow and
Levhari (1969), Kellison (1991, \S4.8), to view
$df(t)=f^\prime(t)\,dt$ as the payment or cash flow in the
interval $[t,t+dt]$ at the density $f^\prime(t)$.

Consequently,
employing the integral (3.4), $f(t)$ is the total cash flow in the
interval $(-\infty,t]$.

\flushpar{\bf Remark 3.1.} Let $\Cal C$ be a finite cash flow
sequence and let $\Cal D$ be the cash flow sequence obtained from
$\Cal C$ by introducing cash flows of 0 at a finite number of
additional partition points. Employing (3.1), it is clear that
$f_{\Cal C} =f_{\Cal D} \in \Cal S$, i.e., the addition of a
finite number of 0 cash flows leaves invariant the corresponding
step function. Conversely, if $f_{\Cal C} =f_{\Cal D}$, then the
cash flow sequences $\Cal C$, $\Cal D$, differ at most by cash
flows of 0 at a finite number of additional partition points.

In view of the above remark, we assume implicitly throughout this
paper that step functions $f_{\Cal C}, f_{\Cal D} \in \Cal S$
satisfy the additional property that the cash flow sequences $\Cal
C, \Cal D$ have a common set of partition points. In particular,
$$
f_{\Cal C} \pm f_{\Cal D}\,=\,f_{\Cal C \pm\Cal D},\quad
\text{where } \Cal C \pm \Cal D =(C_i \pm D_i)_{0\les i\les n}.
$$
For each $\Cal C=(C_i)_{0\le i\le n}$, let $||\,f_{\Cal
C}||=\sup_{0\les p\les n}\{|\,f_{\Cal C}(t_p)|=|\,C_0+\dots
+C_p|\,\}$. Then the step function $f_{\Cal C-\Cal D}\in\Cal S$
satisfies the following estimates.
$$
-||\,f_{\Cal C-\Cal D}||\les
\sum_{i=0}^{i=p}\,(C_i-D_i)\les||\,f_{\Cal C-\Cal D}||\quad 0\les
p\les n. $$ Since
$\sum_{i=p}^{i=q}\,(C_i-D_i)=\sum_{i=0}^{i=q}\,(C_i-D_i)-\sum_{i=0}^{i=p}\,(C_i-D_i)$
for all $p\les q$ it follows that
$$
-2||\,f_{\Cal C-\Cal D}||\les \sum_{i=p}^{i=q}\,(C_i-D_i)\les
2||\,f_{\Cal C-\Cal D}||\quad \text{for all\,}0\les p\les q\les
n.\tag3.5
$$
Since $f_{\Cal C}$ is a step function then also $\|\,f_{\Cal
C}\|=\sup\{|\,f_{\Cal C}(t)|\mid t\in\bold R\}$, i.e., $\|f_{\Cal
C}\|$ is the sup-norm of $f_{\Cal C}$, interpreted in terms of the
sum of the associated cash flows of $f_{\Cal C}\in\Cal S$.

\newpage

 \flushpar{\bf 3.3.
Investment Projects}

\flushpar A regulated payment stream $f\: \bold R \to \bold R$,
minimally supported on the interval $[a,b]$, is an {\it investment
project} if either (i) $f(a) < 0$, or (ii) $f(a) =0$ and there is
a $\delta > 0$ such that the restriction of the function $f$ to
the interval $(a, a+\delta]$ is negative and is non-increasing.
$\Cal I\subset \Cal R$ is the subset of investment projects.
$\IK=\Cal I\cap\RK$ is the subset of investment projects with
minimal support in $K$.

The investment project condition is interpreted to mean that
either $f(a) < 0$ represents the initial outflow (start-up funds)
for the project, or $f(a) =0$ and there is an initial stream of
outflows which constitutes these start-up funds. Employing (3.1),
a step function $f_{\Cal C}\in\Cal S_K$ is an investment project,
i.e., $f_{\Cal C}\in\Cal S_K\cap\,\Cal I_K$, if and only if the
initial cash flow $C_0 < 0$ (an initial outflow).

\vskip .25cm \centerline{\bf 4. Axioms For Balance Functions}

\flushpar{\bf 4.1.} In this section we state the axioms for
balance functions and we prove a classification Theorem 4.5 for
the existence and uniqueness of balance functions. The axioms for
balance functions are stated in terms of the space $\Cal R$ of
regulated payment streams, \S3.2.

A map $\Cal B\rt$ is a {\it balance function}, or {\it future
value function,} if it satisfies the 5 axioms stated below. We
introduce the following preliminary notation.

\item{(i)}$\Cal B(t,f) \equiv\Cal B_t(f) \in \bold R$ is the
balance (future value) of the regulated payment stream $f$ at time
$t\in \bold R$. In financial terms, $\Cal B_t(f)$ is the balance,
or future value, of $f$ at time $t$ due to market forces,
including prevailing interest rates, that act on the payment
stream $f$ over the truncated time interval $(-\infty, t]$, i.e.,
up until the time $t$. \item{(ii)}For each $s\in \bold R$ let $c_s
\in \Cal S$ be the step function payment stream which corresponds
to the single cash flow of 1 at time $s$: $c_s(t) =0$ if $t< s$;
$c_s(t) =1$ if $t\ges s$. For example, let $f_{\Cal C}$ be the
step function payment stream associated to a finite cash flow
sequence $\Cal C=(A_i)_{0\le i\le n}$, as in (3.1) above. Then
$f_{\Cal C}=A_0c_{t_0}+\cdots +A_nc_{t_n}$.

\item{(iii)}A balance function $\Cal B$ induces an ``update map''
$U\equiv U(\Cal B)\: \bold R\times \Cal R \to \Cal R$,
$$
\aligned
U(s,f)(t)\equiv U_s(f)(t)  =\cases 0\quad &\text{if }t< s \\
\Cal B_s(f)c_s +f(t)-f(s) &\text{if }t\ges s.
\endcases \endaligned \tag4.1
$$
For each $s\in \bold R$ the payment stream $U_s(f)\in \Cal R$ has
the property that its cash flow at time $s$ is $\Cal B_s(f)$, the
``updated'' balance at time $s$ of the payment stream $f$ on the
truncated interval $(-\infty,s]$; on the time interval
$(s,\infty)$ the payment streams $U_s(f)$, $f$ coincide:
$U_s(f)(t)-U_s(f)(s)=f(t)-f(s)$. In particular, let $f_{\Cal
C}=A_0c_{t_0}+\cdots +A_nc_{t_n}$ be the step function payment
stream as in (ii) above. Then at each time $t_k$,
$$
U_{t_k}(f_{\Cal C})=\Cal B_{t_k}(f_{\Cal
C})c_{t_k}+A_{k+1}c_{t_{k+1}}+\cdots +A_nc_{t_n}, \quad 0\le k\le
n.
$$
We motivate the updated payment stream, and the replacement axiom
$A_4$ below, in the case of a discrete loan contract, such as a
mortgage contract: the initial debt is $A_0<0$, with $n$ constant
repayments of $R>0$ at the end of each period, calculated at the
rate of interest $i$ per period, i.e., the payment stream
$f=A_0c_0+Rc_1+\cdots +Rc_n$. Classically, the current balance of
the debt after $k$ periods at the rate per period $i$ is,
$$
\Cal B_k(f)=A_0(1+i)^k+R(1+i)^{k-1}+\cdots +R(1+i)+R.
$$
The updated payment stream after $k$ periods is, $U_k(f)=\Cal
B_k(f)c_k+Rc_{k+1}+\cdots +Rc_n$, whose first cash flow is the
current balance $\Cal B_k(f)$ of the debt at time $k$, and whose
remaining $(n-k)$-cash flows are the future unpaid payments of
$R$. After $k$ payment periods, the payment stream $f$ can be
\underbar{replaced} with the updated payment stream $U_k(f)$. The
current balance after $\ell$ periods for the updated payment
stream $U_k(f)$ is, $\Cal B_{\ell}(U_k(f))=\Cal B_{k+\ell}(f)$,
which is easily verified algebraically. This relation ensures the
consistency of calculations of the current balance of the debt,
using either $f$ or the updated payment stream $U_k(f)$, at the
constant rate per period $i$.

\flushpar With these preliminaries, the five axioms for balance
functions are as follows:

\item{$ A_1.$} Let $f\in \Cal R$ and let $g\in \Cal R$ be the
payment stream $g=f+\lambda c_s$ (the addition of a single cash
flow of $\lambda\in\bold R$ at time $s$). For all $r<s$, $\Cal
B_r(g)=\Cal B_r(f)$. Thus cash flows introduced at times later
than $r$ do not contribute to the balance $\Cal B_r(f)$ at time
$r$.\vskip.5cm

\item{$A_2.$}{\it Linearity in the Final Cash Flow:} For each time
$t \in \bold R$, $\Cal B_t(f +\lambda c_t) =\Cal B_t(f) +\lambda$
for all $\lambda\in\bold R$ and all payment streams $f\in\Cal R$.
Informally, market forces in place up until time $t$ do not affect
a cash flow that takes place at the instant $t$. The intended
interpretation of axioms $A_1$, $A_2$ is that the balances $\Cal
B_t(f)$ depend only on the cash flows of the payment stream $f$ on
the time interval $(-\infty,t]$. \vskip .5cm

\item{$A_3.$}{\it Scale:} For all $t\in\bold R, f\in\Cal R$, $\Cal
B_t(\lambda f)=\lambda\Cal B_t(f)$ for all $\lambda\ges 0$.
Furthermore for all $s\les t$, $\Cal B_t(c_s) \ges 0$; $\Cal
B_t(-c_s) \les 0$.

\item{} In particular, for all $t$, $\Cal B_t(\overline{0})=0$,
where $\overline{0}$ is the zero payment stream. Also the balance
at $t\ges s$ of a single cash flow at $s$ does not change sign
(but could be 0). This corresponds to the economic fact that a
single deposit, respectively a single debt, at time $s$ can be
reduced to zero over time but cannot change sign into a debt,
respectively a deposit. Furthermore Axiom $A_3$ states informally
that if all the cash values of a payment stream $f$ are rescaled
by a constant factor $\lambda\ges 0$ then all the future values of
$f$ are rescaled by $\lambda$, i.e., the balance functions are
invariant under a change of monetary unit, a reasonable financial
requirement. We do not assume in general that $\Cal B_t(-f)=-\Cal
B_t(f)$, which is equivalent to the linearity of $\Cal B_t(f)$ in
the payment stream $f\in\Cal R$ (cf. Remark 4.4). However from
$A_3$, $\Cal B_t(uf)=\Cal B_t(-u(-f))=-u\Cal B_t(-f)$ for all
$u\les 0$.

\vskip .5cm \item{$A_4.$}{\it Replacement:} For each $f\in \Cal
R$, $\Cal B_t(f) =\Cal B_t(U_s(f))$ for all $s\les t$ in $\bold
R$. \item{}The Replacement axiom ensures time consistency of
balance functions: For all $s\les t$, the balance $\Cal B_t(f)$ is
equal to the balance at time $t$ of the updated payment stream
$U_s(f)\in\Cal R$ whose cash flow at time $s$ is the balance $\Cal
B_s(f)$ and is such that the payment streams $U_s(f)$, $f$
coincide on $(s,\infty)$. This axiom is motivated by the
discussion above on current balances of a loan contract. \vskip
.5cm \item{$A_5.$}{\it Continuity:} Let $f\in\Cal R_K$ and let $f
=\lim_{n\to \infty} f_n$, where $(f_n)_{n\ges 1}$ is a Cauchy
sequence in $\Cal S_K$ (topology of uniform convergence). For all
$t\in \bold R$, $\Cal B_t(f) =\lim_{n\to \infty} \Cal
B_t(f_n)\in\bold R$.

\flushpar In general a balance map $\Cal B_t(f)$ is
\underbar{non-linear} in $f\in\Cal R$. As explained in (4.2)
below, this non-linearity derives from the difference in general
between ``deposit'' and ``investment'' accumulation functions
discussed in Lemma 4.3. The linear case is discussed in \S4.3 and
also Remark 4.4.

\flushpar{\bf 4.2. Accumulation Functions}

\flushpar Let $\Cal B\rt$ be a balance function. For all $s\les t$
define $a(s,t) =\Cal B_t(c_s)$, the balance at $t$ of a cash flow
of $1$ at $s$. Similarly, for all $s\les t$ define $b(s,t) =-\Cal
B_t(-c_s)$, the negative of the balance at $t$ of a cash flow of
$-1$ at $s$. Employing axiom $A_3$, $a(s,t)\ges 0$, $b(s,t)\ges 0$
for all $s\les t$. \proclaim{Lemma 4.1}$a(s,t)$, $b(s,t)$ are
accumulation functions, called the deposit, respectively the
investment accumulation function for the balance map $\Cal B\rt$.
\endproclaim

\flushpar{\bf Proof.} Let $r\les s\les t$. We prove that $a(s,t)$,
$b(s,t)$ satisfy the multiplicative property (2.1) for
accumulation functions. Employing the replacement axiom $A_4$,
$a(r,t) =\Cal B_t(c_r) =\Cal B_t(U_s(c_r))$. From (4.1) the
payment stream $U_s(c_r)(u)$ is 0 for all $u< s$ and is the
constant $\Cal B_s(c_r)\ges 0$ for all $u\ges s$. Hence $U_s(c_r)
=\lambda c_s$, $\lambda=\Cal B_s(c_r)\ges 0$. Consequently,
$$
\aligned
a(r,t) &=\Cal B_t(U_s(c_r)) =\Cal B_t(\Cal B_s(c_r)\cdot c_s) \\
&=\Cal B_s(c_r)\cdot \Cal B_t(c_s)\quad \text{by $A_3$}.\\
\text{Hence }a(r,t) &=a(r,s)\cdot a(s,t).
\endaligned
$$
Similarly, employing the replacement axiom $A_4$, $b(r,t) =-\Cal
B_t(U_s(-c_r))$. Employing (4.1), the payment stream
$U_s(-c_r)(u)=0$ for all $u<s$ and is the constant $\Cal
B_s(-c_r)\les 0$ for all $u\ges s$. Hence $U_s(-c_r) =\lambda
c_s$, $\lambda=\Cal B_s(-c_r)\les 0$. Consequently,
$$
\aligned
b(r,t) &=-\Cal B_t(U_s(-c_r)) =-\Cal B_t(\Cal B_s(-c_r)\cdot c_s) \\
&= +\Cal B_s(-c_r)\cdot \Cal B_t(-c_s) \quad \text{by $A_3$}. \\
\text{Hence }b(r,t) &=b(r,s)\cdot b(s,t).
\endaligned
$$
Thus $a(s,t)$, $b(s,t)$ both satisfy the multiplicative property
for accumulation functions. Furthermore, for all $t\in \bold R$,
$a(t,t)=b(t,t)=1$. Indeed, employing axiom $A_2$, for all
$t\in\bold R$,
$$
a(t,t)=\Cal B_t(c_t)=\Cal B_t(\overline{0}+1c_t)=\Cal
B_t(\overline{0})+1=1.
$$
Similarly employing $A_2$, $\Cal B_t(-c_t)=\Cal
B_t(\overline{0}-c_t)=\Cal B_t(\overline{0})-1=-1$. Hence for all
$t\in\bold R$, $b(t,t)= -\Cal B_t(-c_t)=1$, which completes the
proof of the lemma. \qed

\proclaim{Lemma 4.2}Let $u\in \bold R$ and suppose $f\in \Cal R$
satisfies $f(t) =0$ for all $t< u$. Then $\Cal B_u(f) =f(u)$.
\endproclaim

\flushpar{\bf Proof.} One may suppose $f=\lim_{n\to\infty}f_n$,
where $(f_n)$ is a sequence of step function payment streams such
that $f_n(t)=0$ for all $t<u$. If
$f_n(t)=\Sigma_{i=0}^mA^n_ic_{t_i}$, $t_0=u$, then by Axioms
$A_1,A_2$, $\Cal B_u(f_n)=\Cal B_u(A^n_0c_u)=A^n_0=f_n(u)$.
Employing axiom $A_5$ it follows that $\Cal
B_u(f)=\lim_{n\to\infty}\Cal
B_u(f_n)=\lim_{n\to\infty}f_n(u)=f(u)$\qed

\proclaim{Lemma 4.3. The Basic Computation}Let $r\les s$ and
suppose $f =xc_r +yc_s$ (thus $f$ represents a cash flow of $x$ at
$r$ and a cash flow of $y$ at $s$). Then,
$$
\aligned
\Cal B_s(f) =\cases xa(r,s) +y \quad &\text{if }x\ges 0 \\
xb(r,s) +y &\text{if }x\les 0.
\endcases \endaligned
$$
\endproclaim

\flushpar{\bf Proof.} Employing axiom $A_2$, $\Cal B_s(f) =\Cal
B_s(xc_r +yc_s) =\Cal B_s(xc_r) +y$. In case $x\ges 0$, employing
the Scale axiom $A_3$, $\Cal B_s(xc_r) =x\Cal B_s(c_r) = xa(r,s)$.
In case $x\les 0$, employing axiom $A_3$, $\Cal B_s(xc_r) = -x\Cal
B_s(-c_r) =xb(r,s)$. These two cases prove the lemma. \qed

We remark that Lemma 4.3 shows that the balance $\Cal B_s(f)$ is
governed by the deposit accumulation function $a(r,s)$ in case the
previous balance $\Cal B_r(f)=x\ges 0$, or by the investment
accumulation function $b(r,s)$ in case the previous balance $\Cal
B_r(f)=x\les 0$. This distinction between deposit and investment
accumulation functions derives from our axioms. In financial
terms, the investment accumulation function applies to the current
debt, and the deposit accumulation function applies to the current
surplus, a distinction first employed by Teichroew et al., (1965a)
in their study of IRR functions.

Employing Lemma 4.3, we show below that the balance map, when
restricted to step function payment streams, $\Cal B\: \bold
R\times \Cal S\to \bold R$, is uniquely determined by the deposit
and investment accumulation functions $a(s,t), b(s,t)$. Uniqueness
of balance functions, $\Cal B\rt$, then follows from the
continuity axiom $A_5$. The existence of balance functions
that satisfy all of the axioms is proved in Theorem 4.5.

\flushpar Let $f=D_0c_{t_0}+\cdots +D_nc_{t_n}\in\Cal S$ be a step
function payment stream whose successive cash flows $D_j$ occur at
times $t_j$, $0\le j\le n$. Let $\Cal B_j(f)\equiv \Cal
B_{t_j}(f)$ denote the current balance at time $t_j$, $0\le j\le
n$. Employing (4.1), the updated payment stream at time $t_k$ is
$U_k(f)=\Cal B_k(f)c_k+D_{k+1}c_{k+1}+\cdots +D_nc_n$, $0\le k\le
n$. Applying Axiom $A_1$ and the replacement axiom $A_4$ it
follows that the balance at time $t_{j+1}$ is,
$$
\aligned \Cal B_{j+1}(f)&=\Cal B_{j+1}(U_j(f))=\Cal
B_{j+1}\left(\Cal
B_j(f)c_{t_j}+D_{j+1}c_{t_{j+1}}+\cdots+D_nc_{t_n}\right)\\
&=\Cal B_{j+1}(\Cal B_j(f)c_{t_j}+D_{j+1}c_{j+1})\quad 0\le j\le
n-1.
\endaligned
$$
Applying Lemma 4.3 to the times $t_j$, $t_{j+1}$, and to the step
function payment stream (two cash flows), $\Cal
B_j(f)c_{t_j}+D_{j+1}c_{j+1}$, it follows that
$$
\aligned \bc1  =\cases a(t_j ,t_{j +1})\,\Cal B_j(f)  +D_{j +1}\
&\text{if }
\Cal B_j(f) \ges 0, \\
b(t_j,t_{j +1})\,\Cal B_j(f) +D_{j +1}\ &\text{if }\Cal B_j(f)
\les 0,
\endcases\quad 0\le j\le n-1.
\endaligned \tag4.2
$$
Thus on the interval $[t_j ,t_{j +1}]$, the balance $\Cal B_j(f)$
at time $t_j$ accumulates with respect to the accumulation
function $a(s,t)$ in case the balance $\Cal B_j(f)\ges 0$, or with
respect to the accumulation function $b(s,t)$ in case the balance
$\Cal B_j(f)\les 0$. Only the cash flows $D_0,\dots ,D_j$ of $f$
enter into the computation of $\Cal B_j(f)$. The iteration scheme
(4.2) is non-linear in the payment stream $f\in\Cal S$; in general
there is no closed form expression for the balances $\Cal B_j(f)$.
This type of iteration scheme for balance functions was first
considered by Teichroew et al., (1965a,b), in the special case
that $a(s,t) =d^{t -s}$, $b(s,t) =x^{t -s}$, where $d> 0$ is a
constant ``deposit'' compounding factor and $x > 0$ is a constant
``investment'' compounding factor; in addition, these authors
assume a constant period, i.e., the intervals $[t_i ,t_{i -1}]$
have equal length, $1\les i\les n$. These balances $\Cal
B_j(f)\equiv\Cal B_j(f)(a,b)$ are therefore designated throughout
this paper as T.R.M. balances, with respect to the deposit and
investment accumulation functions $a(s,t)$, $b(s,t)$.

\vskip.25cm \flushpar{\bf Remark 4.4.} A special case of interest
for T.R.M. balances occurs in the case $a =b$, i.e., the deposit
and investment accumulation functions are equal. In this case the
iteration scheme (4.2) simplifies:
$$
\bc1  = a(t_j ,t_{j +1})\Cal B_j(f) + D_{j +1}.\tag4.3
$$
From (2.1), (4.3) one obtains closed form expressions for the
successive T.R.M. balances:
$$
\Cal B_j(f) =D_0 a(t_0, t_j) + D_1 a(t_1 ,t_j) + \cdots + D_j,
\quad 0\les j\les n.\tag4.4
$$
\flushpar Thus in case $a=b$ the balances $\Cal B_j(f)$ are
\underbar{linear} in the payment stream $f\in\Cal S$; hence $\Cal
B_t(\lambda f)=\lambda\Cal B_t(f)$ for all $\lambda\in\bold R$. If
$a(t,s) =(1 +i)^{t -s}$ ($i \ges -1$), at constant rate per period
$i$ (the time period in 1 unit), then one recovers the classical
balance (future value) at the end of the project ($n$ periods)
$$
\Cal B_n (f)\,=\,\sum_{k =0}^{k =n} D_k (1 +i)^{n -k}. \tag4.5
$$
Thus the T.R.M. balances (4.2) include, as a special case, the
classical future value calculations in financial mathematics with
respect to a constant rate $i$ per period. \vskip.25cm

\flushpar{\bf 4.2.} In this section we prove that balance
functions exist, subject to some mild restrictions on the deposit
and investment accumulation functions.\vskip.25cm
\proclaim{Theorem 4.5}Let $a\ges 0$, $b\ges 0$ be accumulation
functions which are bounded above by a monotone increasing
accumulation function $y(s,t)$: $a\les y$, $b\les y$. There is a
unique balance function $\Cal B\rt$ whose corresponding deposit
and investment accumulation functions are respectively $a(s,t)$,
$b(s,t)$.
\endproclaim

\flushpar{\bf Proof.} Note that the hypothesis of the theorem is
satisfied if both $a,b$ are positive and of bounded variation on
all compact intervals $I\subset\bold R$ (cf. \S2.2). For arbitrary
accumulation functions $a(s,t), b(s,t)$ the iteration scheme (4.2)
for T.R.M. balances defines a balance function, $\Cal B\: \bold
R\times\Cal S\to\bold R$, on the subset of step function payment
streams. Clearly these T.R.M. balances satisfy axioms $A_1$,
$A_2$, $A_3$, applied to payment streams $f\in\Cal S$.
Furthermore, with respect to these T.R.M. balances, it follows
from (4.1) that the update map $U\: \bold R\times\Cal S\to\Cal S$.
Indeed, $U_s(f)\in\Cal S$ is a step function payment stream whose
first cash flow is the balance $\Cal B_s(f)$, itself defined by
iteration as in (4.2), and such that $U_s(f)$,$f$ have the same
cash flow sequence in $(s,\infty)$. Consequently continuing the
iteration scheme (4.2) for all $t\ges s$, the replacement axiom
$A_4$ is satisfied for all $f\in\Cal S$. To complete the existence
proof we extend this balance map $\Cal B\: \bold R\times\Cal
S\to\bold R$ to a balance map on all regulated payment streams
$f\in\Cal R$, based on the limit process in axiom $A_5$. Thus if
$f=\lim_{n\to \infty}f_n\in \Cal R_K$ (topology of uniform
convergence) where for all $n$, $f_n\in \Cal S_K$ is a step
function payment stream, then the analytic problem is to prove
that, for all $t$, the sequence of balances $\left(\Cal
B_t(f_n)\right)_{n\ge 1}$ is Cauchy, hence $\lim_{n\to\infty}\Cal
B_t(f_n)$ exists. It is here that the hypothesis $a\les y$, $b\les
y$, is employed to establish the estimates needed to carry out the
limiting process defined by axiom $A_5$. The key estimate is
Proposition 4.8.

To emphasize the dependence on the accumulation functions
$a(s,t),b(s,t)$ the balance map $\Cal B\:\bold R\times\Cal
S\to\bold R$ will be written $\Cal B_t(f)(a,b)$, or $\Cal
B_t(f_{\Cal C})(a,b)$, to include also the dependence on the cash
flow sequence $\Cal C =(C_j)_{0\les j\les n}$.
\proclaim{Lemma
4.6}Let $a(s,t)$, $b(s,t)$, $c(s,t)$, $d(s,t)$ be accumulation
functions such that $a\les c$, $b\les d$. For all step functions
$f_{\Cal C} \in \Cal S$ ($\Cal B_j(f_{\Cal C})\equiv\Cal
B_{t_j}(f_{\Cal C}))$,
$$ \Cal B_j
(f_{\Cal C})(a,b)\,\les \,\Cal B_j(f_{\Cal C})(c,b); \quad \Cal
B_j(f_{\Cal C})(a,b)\, \ges \,\Cal B_j(f_{\Cal C})(a,d), \quad
0\les j\les n.
$$
\endproclaim
\flushpar{\bf Proof.} The intuitive financial content of the lemma
may be expressed as follows, and is the main idea underlying
Teichroew et al.,\,(1965a). $\Cal B_j (f_{\Cal C})(a,b)$ is the
balance at time $t_j$ in a financial account which credits
interest in the case of positive balances according to the deposit
accumulation function $a(s,t)$ and which charges interest in the
case of negative balances (overdrafts) according to the investment
(for the financial institution) accumulation function $b(s,t)$.
Evidently, for a given cash flow sequence $\Cal C$, the balance at
time $t_j$ increases if positive balances at previous times are
credited interest at a higher rate, and decreases if overdrafts at
previous times are charged interest at a higher rate.

Formally, the proof is by induction. Assuming $\Cal B_j (f_{\Cal
C}) (a,b)\les \jc (c,b)$, $\jc(a,b)\ges \jc(a,d)$ (note that $\Cal
B_0 (f_{\Cal C}) =C_0$ for all choices of accumulation functions),
the inductive step is proved from (4.2), taking into account the
sign of the balance $\jc$ at time $j$. The details are trivial and
are left to the reader. \qed

\proclaim{Corollary 4.7}Let $a(s,t),b(s,t)$ be accumulation
functions and suppose $y(s,t)$ is an accumulation function which
is a common upper bound: $a\les y$, $b\les y$. For all step
functions $f_{\Cal C} \in \SK$ $($\,$0$ is the zero accumulation
function$)$,
$$
\Cal B_j(f_{\Cal C})(0,y)\,\les \,\Cal B_j(f_{\Cal C})(a,b)\,\les
\,\Cal B_j(f_{\Cal C})(y,0),\quad 0\les j\les n.
$$
\endproclaim
\proclaim{Proposition 4.8}Let $f_{\Cal C},f_{\Cal D}\in\Cal S$ be
step function payment streams; $\Cal C=(C_j)_{0\le j\le n}$, $\Cal
D=(D_j)_{0\le j\le n}$. Let $a\ges 0$, $b\ges 0$ be accumulation
functions that are bounded above by a monotone increasing
accumulation function $y(s,t)$: $a\les y$, $b\les y$. Then,
$$
|\,\Cal B_j(f_{\Cal C})(a,b)-\Cal B_j(f_{\Cal D})(a,b)|\les
2y([t_0,t_j])\cdot||\,f_{\Cal C-\Cal D}||,\quad 0\les j\les n.
$$
\endproclaim
\flushpar{\bf Proof.} Employing Remark 3.1 we assume that the cash
flows of $\Cal C$, $\Cal D$ occur at a common set of partition
points, $t_0 < t_1 < \cdots < t_n$. The Proposition is proved by
induction, based on the following two lemmas and the iteration
scheme (4.2).

\proclaim{Lemma 4.9}$\jc (a,b) -\jd (a,b) \les \Cal B_j (f_{\Cal C
-\Cal D})(y,0)$, $0\les j\les n$ (the index $j$ indicates the
balance at time $t_j$).
\endproclaim

\proclaim{Lemma 4.10}$\jc (a,b) -\jd (a,b) \ges \Cal B_j (f_{\Cal
C -\Cal D})(0,y)$, $0\les j\les n$.
\endproclaim
\flushpar Note that the lemmas are both true with equality at the
index $j =0$ (for all accumulation functions) since $\Cal
B_0(f_{\Cal C})(\cdot,\cdot)=C_0$; $\Cal B_0(f_{\Cal
D})(\cdot,\cdot)=D_0$; $\Cal B_0(f_{\Cal C-\Cal
D})(\cdot,\cdot)=C_0-D_0$.

\flushpar{\bf Proof of Lemma 4.9.} Let $\dj =\Cal B_j (f_{\Cal
C-\Cal D})(y,0) -(\jc (a,b) - \jd (a,b))$, $0\les j\les n$.
Inductively, we assume $\dj \ges 0$ and we prove $\cd \ges 0$.
There are four cases, depending on the signs of $\jc (a,b)$, $\jd
(a,b)$. Employing the iteration scheme (4.2), $\cd$ is computed
from $\dj$ by calculating the change in the balance functions over
the interval $[t_j,t_{j +1}]$. The occurrences of the cash flows
$C_{j +1},D_{j +1}$ at time $t_{j +1}$ in $\cd$ cancel out, hence
are omitted in the computations below for $\cd$. For notational
convenience let $\jc =\jc(a,b)$, $\jd =\jd(a,b)$, $\Cal
B_j(f_{\Cal C -\Cal D}) =\gy$. \flushpar {\smc Case 1}: $\jc \ges
0$, $\jd \ges 0$.
$$
\aligned
\cd &= \sup \{0,\Cal B_j(f_{\Cal C -\Cal D})\yt \} -(\jc   -\jd )\at \\
&\ges \Cal B_j(f_{\Cal C -\Cal D})\at -(\jc  -\jd )\at \\
&= \dj\, \at \ges 0.
\endaligned
$$
\flushpar {\smc Case 2}: $\jc  \ges 0$, $\jd  \les 0$.
$$
\aligned
\cd &= \sup\{ 0, \Cal B_j (f_{\Cal C -\Cal D}) \yt \} -\jc \at +\jd \bt \\
&\ges \Cal B_j (f_{\Cal C -\Cal D}) \yt -\jc  \yt + \jd \yt \\
&= \dj \, \yt \ges 0.
\endaligned
$$
\flushpar {\smc Case 3}: $\jc  \les 0$, $\jd  \ges 0$.
$$
\aligned
\cd &= \sup\{ 0, \Cal B_j (f_{\Cal C -\Cal D})\yt \} -\jc \bt +\jd \at \\
& \ges 0  \quad \text{(each term is $\ges 0$)}.
\endaligned
$$
\flushpar {\smc Case 4}: $\jc  \les 0$, $\jd  \les 0$.
$$
\aligned
\cd &= \sup \{0,\Cal B_j (f_{\Cal C -\Cal D}) \yt \} -(\jc   -\jd )\bt \\
&\ges \Cal B_j (f_{\Cal C -\Cal D}) \bt -(\jc  -\jd )\bt \\
&= \dj\, \bt \ges 0.
\endaligned
$$
The above four cases prove the inductive step and hence the lemma
is proved. \qed \

\flushpar{\bf Proof of Lemma 4.10.} Let $\dj =\Cal B_j (f_{\Cal
C-\Cal D})(0,y) -(\jc (a,b) - \jd (a,b))$. Inductively, employing
the iteration scheme (4,2), we assume $\dj \les 0$ and we prove
$\cd \les 0$. Again, the occurrences of the cash flows $C_{j +1},
D_{j +1}$ at time $t_{j +1}$ in $\cd$ cancel out, hence are
omitted. For notational convenience let $\jc = \jc(a,b)$, $\jd
=\jd(a,b)$, $\Cal B_j (f_{\Cal C -\Cal D}) =\ly$.

\flushpar {\smc Case 1}: $\jc  \ges 0$, $\jd  \ges 0$.
$$
\aligned
\cd &= \inf \{0,\Cal B_j(f_{\Cal C -\Cal D}) \yt \} -(\jc   -\jd )\at \\
&\les \Cal B_j (f_{\Cal C -\Cal D})\at -(\jc  -\jd )\at \\
&= \dj\, \at \les 0.
\endaligned
$$
\flushpar {\smc Case 2}: $\jc  \ges 0$, $\jd  \les 0$.
$$
\aligned
\cd &= \inf\{ 0, \Cal B_j (f_{\Cal C -\Cal D}) \yt \} -\jc \at +\jd \bt \\
&\les 0 \quad \text{(each term is $\les 0$)}.
\endaligned
$$
\flushpar {\smc Case 3}: $\jc  \les 0$, $\jd \ges 0$.
$$
\aligned
\cd &= \inf\{ 0, \Cal B_j(f_{\Cal C -\Cal D}) \yt \} -\jc \bt +\jd \at \\
&\les \Cal B_j(f_{\Cal C -\Cal D}) \yt -\jc  \yt + \jd  \yt \\
&= \dj\,\yt \les 0.
\endaligned
$$
\flushpar {\smc Case 4}: $\jc  \les 0$, $\jd  \les 0$.
$$
\aligned
\cd &= \inf \{0,\Cal B_j(f_{\Cal C -\Cal D})\yt \} -(\jc   -\jd )\bt \\
&\les \Cal B_j(f_{\Cal C -\Cal D}) \bt -(\jc  -\jd )\bt \\
&= \dj\, \bt \les 0.
\endaligned
$$
The above four cases prove the inductive step and hence the lemma
is proved. \qed

\flushpar Returning to the proof of Proposition 4.8, for all $j$,
$0\les j\les n$, it follows from Lemma 4.9, Lemma 4.10 that,
$$
\Cal B_j(f_{\Cal C -\Cal D})(0,y) \les \Cal B_j (f_{\Cal C})(a,b)
-\Cal B_j (f_{\Cal D})(a,b) \les \Cal B_j(f_{\Cal C -\Cal
D})(y,0).\tag4.6
$$
We now estimate the end terms of (4.6). Employing the iteration
scheme (4.2), note that if the deposit, respectively investment,
accumulation function is 0 then a balance $\Cal B_p(f_{\Cal C-\Cal
D})(y,0)\les 0$ at time $t_p$ implies that the next balance $\Cal
B_{p+1}(f_{\Cal C-\Cal D})(y,0)= C_{p+1}-D_{p+1}$ at time
$t_{p+1}$, respectively a balance $\Cal B_p(f_{\Cal C-\Cal
D})(0,y)\ges 0$ at time $t_p$ implies that the next balance $\Cal
B_{p+1}(f_{\Cal C-\Cal D})(0,y)= C_{p+1}-D_{p+1}$ at time
$t_{p+1}$.

There is a largest $k$, $1\les k\les j$, such that $\Cal
B_{k-1}(f_{\Cal C-\Cal D})(y,0)\les 0$\,; hence $\Cal B_k(f_{\Cal
C-\Cal D})(y,0)=C_k-D_k$, and if $k<j$, then $\Cal B_r(f_{\Cal
C-\Cal D})\ges 0$, $k\les r<j$. Consequently, employing the
iteration scheme (4.2), one computes the balance at time $t_j$
$$
\Cal B_j (f_{\Cal C -\Cal D})(y,0) =(C_k -D_k) y(t_k,t_j) +(C_{k
+1} -D_{k +1}) y(t_{k +1},t_j) + \cdots + (C_j -D_j). \tag4.7
$$
Since the sequence $(y(t_i,t_j))_{i\les j}$ is monotone
decreasing, it follows from the classical Abel's lemma for finite
series, cf. Spivak (1980, p.\,368), that
$$
\Cal B_j(f_{\Cal C -\Cal D})(y,0)\les y(t_k,t_j)\cdot\Sigma\tag4.8
$$
where $\Sigma =\sup\{(C_k -D_k) +\cdots +(C_p -D_p)\mid k\les
p\les j\}$. Employing (3.5), it follows that $|\,\Sigma| \les
2\,\|f_{\Cal C-\Cal D}\|$. Since $y(s,t)$ is monotone increasing,
$y(t_k,t_j)\le y(t_0,t_j)$. Hence
$$
\Cal B_j(f_{\Cal C -\Cal D})(y,0) \les 2\,y(t_0,t_j)\cdot\|f_{\Cal
C-\Cal D}\|. \tag4.9
$$
Similarly, for the other end term $\Cal B_j(f_{\Cal C -\Cal
D})(0,y)$ of (4.6), there is a largest $k$ such that $\Cal
B_{k-1}(f_{\Cal C-\Cal D})(0,y)\ges 0$\,; hence $\Cal B_k(f_{\Cal
C-\Cal D})(0,y)=C_k-D_k$, and if $k<j$, $\Cal B_r(f_{\Cal C-\Cal
D})\les 0$, $k\les r<j$. Consequently, employing the iteration
scheme (4.2), one computes the balance at time $t_j$
$$
\Cal B_j (f_{\Cal C -\Cal D})(0,y) =(C_k -D_k) y(t_k,t_j) +(C_{k
+1} -D_{k +1}) y(t_{k +1},t_j) + \cdots + (C_j -D_j). \tag4.10
$$
Since the sequence $(y(t_i,t_j))_{i\les j}$ is monotone
decreasing, it follows from Abel's lemma that,
$$
\Cal B_j(f_{\Cal C -\Cal D})(0,y) \ges y(t_k,t_j)
\cdot\sigma,\tag4.11
$$
where $\sigma =\inf\{(C_k -D_k) +\cdots +(C_p-D_p)\mid k\les p\les
j\}$. Employing (3.5), it follows that $|\,\sigma|\ges
-2\,\|f_{\Cal C-\Cal D}\|$. Since $y(s,t)$ is monotone increasing,
$y(t_k,t_j)\le y(t_0,t_j)$. Hence
$$
\Cal B_j(f_{\Cal C -\Cal D})(0,y) \ges -2\,y(t_0,t_j)\cdot
\|\,f_{\Cal C-\Cal D}\|. \tag4.12
$$
Employing (4.6) and the estimates (4.9), (4.12), one obtains the
inequality,
$$
-2\,y(t_0,t_j)\cdot \|\,f_{\Cal C -\Cal D}\| \les \Cal B_j(f_{\Cal
C})(a,b) -\Cal B_j(f_{\Cal D})(a,b) \les
2\,y(t_0,t_j)\cdot\|\,f_{\Cal C -\Cal D}\|\quad 0\les j\les n.
\tag4.13
$$
Consequently, $|\,\Cal B_j(f_{\Cal C})(a,b) -\Cal B_j(f_{\Cal
D})(a,b)| \les 2\,y(t_0,t_j)\cdot\|\,f_{\Cal C-\Cal D}\|$, $0\les
j\les n$, which completes the proof of Proposition 4.8. \qed

Returning to the proof of Theorem 4.5, let $f\in\RK$ and let
$(f_n=f_{\Cal C_n})_{n\ges 1}$ be a Cauchy sequence of step
function payment streams in $\SK$ such that
$f=\lim_{n\to\infty}f_n$. From (4.13)
$$
|\,\Cal B_j(f_n)(a,b)-\Cal B_j(f_m)(a,b)|\les
2\,y(t_0,t_j)\cdot\|\,f_n-f_m\|,\quad 0\les j\les n.\tag4.14
$$
From the remarks following (3.5), $\|\,f_n-f_m\|$ is the sup-norm.
Since $\SK$ is dense in $\RK$ in the sup-norm topology, it follows
from (4.14) that at time $t$ ($t=t_j$ for some $j$) the sequence
of balances $(\Cal B_t(f_m)\in\bold R)_{m\ges 1}$ is Cauchy. Hence
there is a balance map $\Cal B\rt$ such that $\Cal
B_t(f)(a,b)=\lim_{m\to\infty}\Cal B_t(f_m)(a,b)$ for all
$t\in\bold R$. Since the balance map $\Cal B\:\bold R\times \Cal
S\to\bold R$ satisfies all of the axioms $A_i$, $1\les i\les 4$,
it follows in the limit that the balance map $\Cal B\rt$ satisfies
these axioms and, by construction, also the continuity axiom
$A_5$, which proves Theorem 4.5.\qed

\vskip .25cm \flushpar{\bf 4.3. Linear Balance maps.} We consider
now the special case that $a=b$ in Theorem 4.5, i.e., the deposit
and investment accumulation functions are equal. Employing (4.4)
the condition $a=b$ is equivalent to an additional axiom that a
balance map $\Cal B\:\bold R\times \Cal S\to\bold R$ is {\it
linear} in the payment streams $f\in\Cal S$; hence by the
continuity axiom $A_5$, the balance map $\Cal B\rt$ is linear in
the regulated payment stream $f\in\Cal R$. Suppose
$a(s,t)=b(s,t)=e^{\,g(t)-g(s)}$,where $g(t)$ is of bounded
variation on all compact intervals. Applying Theorem 4.5, for each
$f\in\Cal R_K$, $K=[c,d]$, we write the corresponding balance map
$$\Cal
B_t(f)=\int_{-\infty}^{\,t} e^{\,g(t)-g(s)}df(s).\tag4.15
$$
This ``generalized''integral (4.15) is linear in $f\in\RK$ and
coincides with the classical balance map (future value map) in
case $f\in\RK$ is also of bounded variation, i.e., (4.15)
specializes to a Lebesgue-Stieltjes integral in this case. Thus
(4.15) reduces to the future value calculations (4.4), (4.5), if
$f=f_{\Cal C}\in\SK$ is a step function payment stream, and
$a(s,t)=(1+i)^{t-s}$.

In this respect, Theorem 4.5 generalizes the work of Norberg
(1990), Promislow (1994), who propose axioms, including a
linearity axiom, for balance functions, denoted by these authors
as {\it valuation} functions, on payment streams that in their
theory are Borel measures $\mu$ on $\bold R$. Thus $\mu([a,b])$ is
the total cash flow in the interval $[a,b]\subset \bold R$. In
Norberg (1990), payment measures are non-negative; Promislow
(1994) generalizes this measure-theoretic approach to include
payment streams that are signed Borel measures on $\bold R$, i.e.,
that reflect transactions that may have both positive and negative
payments. Since these Borel measures can be represented by
functions of bounded variation on compact intervals it follows
that the valuation functions of these authors, when calculated at
the time of the final cash flow of the payment stream, is a
special case of the integral (4.15) where $f\in\Cal R_K$ is of
bounded variation. \vskip .25cm \centerline{\bf 5. Internal Rate
of Return}

\flushpar{\bf 5.1.} Let $a(s,t)$ be a positive accumulation
function of bounded variation. As explained in \S2.2 there is a
positive monotone increasing accumulation function $y(s,t)$,
respectively a positive monotone decreasing accumulation function
$b(s,t)$ such that $0<a\les y$, respectively $a\ges b>0$.

The accumulation function $x(s,t)=x^{t-s}$, $t\les s$, is positive
if $x>0$ and is the zero accumulation function if $x=0$. Let
$x_1=\sup\{x,1\}$, $x_2=\inf\{x,1\}$. The accumulation function
$x_1^{t-s}$, respectively $x_2^{t-s}$, is monotone increasing,
respectively monotone decreasing. Consequently, the product
accumulation function $z(s,t)=y(s,t)\cdot x_1^{t-s}$ is a positive
monotone increasing common upper bound, respectively if $x>0$,
$c(s,t)=b(s,t)\cdot x_2^{t-s}$ is a positive monotone decreasing
lower bound: for all $s\les t$ (cf. \S2.2)
$$ a(s,t)\les z(s,t),\quad x^{t-s}\les z(s,t)\,;\quad 0< c(s,t)\les a(s,t),
\quad 0<c(s,t)\les x^{t-s}.\tag5.1
$$
In what follows we let $a>0$ be a fixed deposit accumulation
function of bounded variation and we let $x^{t-s}$, $x\ges 0$, be
a variable investment accumulation function. With respect to these
accumulation functions, and also the common upper bound $z(s,t)$
in (5.1), it follows from Theorem 4.5 that there is a unique
balance map $\Cal B\rt$ that satisfies all of the axioms $ A_i$,
$1\les i\les 5$. Throughout we employ the simplified notation
$\Cal B_t^{\,x}(f)=\Cal B_t(f)(a,x)$, $f\in\Cal R$, to indicate
the dependence on the variable investment accumulation function
$x^{t-s}$.

Let $\Cal D\subset C^0([0,\infty),\bold R)$, in the compact-open
topology, be the subspace of continuous functions
$f\:[0,\infty)\to\bold R$ such that either $f$ is strictly
decreasing with a unique root $f(x)=0$, or $f$ is negative and
non-increasing (for example a constant function $<0$). Let
$\nu\:\Cal D\to[0,\infty)$ be the Lebesgue measure,
$\nu(f)=m(X_f)$,  $X_f=\{x\in[0,\infty)\mid f(x)\ges 0\}$. In
particular $\nu(f)=0$ if $f$ is negative, and $\nu(f)=x_0$ if
$f(x_0)=0$. Promislow and Spring (1996, Theorem 4.3) prove that
the measure $\nu\:\Cal D\to [0,\infty)$ is continuous.

Let $g=g_{\Cal C}\in\Cal S\cap\Cal I$ be a step function payment
stream that is an investment project: $\Cal C=(C_i)_{0\les i\les
n}$ is a finite cash flow sequence such that $C_0<0$ since
$g\in\Cal I$. The central point, proved in the next lemma, is that
for each $t\ges t_0$, the function $\Cal B^{\,x}_t(g)$, as a
function of $x$, lies in $\Cal D$. For example, at the initial
time $t_0$, the balance function $\Cal B^{\,x}_{t_0}(g)=C_0<0$ (a
constant function of $x$), hence $\Cal B^{\,x}_{t_0}(g)\in\Cal D$.
The IRR of investment projects will be defined in \S5.2 in terms
of the measure $\nu(f)$ above on the space $\Cal D$.
\proclaim{Lemma 5.1}As a function of the variable
$x\in[0,\infty)$, for each $t\ges t_0$ the balance function $\Cal
B^{\,x}_t:\Cal S\cap\,\Cal I\to\Cal D$. If $t>t_0$ then for each
$g\in\Cal S\cap\,\Cal I$ the function $\Cal B^{\,x}_t(g)\in\Cal D$
is a continuous, strictly decreasing function of $x\in[0,\infty)$
such that $\lim_{x\to\infty}\Cal B^{\,x}_t(g)=-\infty$.
\endproclaim

\flushpar{\bf Proof.} let $\Cal B^{\,x}_j(g)\equiv\Cal
B^{\,x}_{t_j}(g)$, where $t_j$ is the time of the cash flow $C_j$,
$0\les j\les n$. By introducing a cash flow of 0 at $t$ if
necessary, we may assume that $t=t_j$ for some $j\ges 1$. As noted
above $\Cal B^{\,x}_0(g)\in\Cal D$ is the constant function
$C_0<0$. Inductively on $j$, suppose $\Cal B^{\,x}_j(g)\in\Cal D$.
Let $r_j=\nu(\Cal B^{\,x}_j(g))\in[0,\infty)$. Thus if $r_j>0$
then $\Cal B^{r_j}_j(g)=0$; $\Cal B^{\,x}_j(g)>0$ if $x<r_j$;
$\Cal B^{\,x}_j(g)< 0$ if $x>r_j$. Applying (4.2) with respect to
the deposit accumulation function $a>0$ and investment
accumulation function $x^{t-s}$,
$$
\Cal B^{\,x}_{j+1}(g)=\cases a(t_j,t_{j+1})\Cal B^{\,x}_j(g)+C_{j+1}\quad &\text{if }x\in[0,r_j)\ (\Cal B_j^{\,x}(g)>0)\\
x^{t_{j+1}-t_j}\Cal B^{\,x}_j(g)+C_{j+1}\quad &\text{if }x\ges
r_j\ (\Cal B_j^{\,x}(g)\le0).\endcases\tag5.2
$$
If $r_j=0$ then only the second alternative in (5.2) applies. One
easily checks that $\Cal B^{\,x}_{j+1}(g)$ is a continuous
function of $x$ (if $r_j>0$ then $\Cal B^{\,r_j}_j(g)=0$).
Applying (5.2) at $j=0$ ($C_0<0$; $r_0=0$), $\Cal
B^{\,x}_1(g)=C_0x^{t_1-t_0}+C_1$ for all $x\ges 0$; hence $\Cal
B_1(\Cal C)\in\Cal D$ is a strictly decreasing function such that
$\lim_{x\to\infty}\Cal B^{\,x}_1(g)=-\infty$. Suppose inductively
in addition that $\Cal B^{\,x}_j(g)\in\Cal D$ is a strictly
decreasing function of $x$ such that $\lim_{x\to\infty}\Cal
B^{\,x}_j(g)=-\infty$, $j\ges 1$. Since $a>0$, employing (5.2), if
$r_j>0$ then $\Cal B^{\,x}_{j+1}(g)$ is strictly decreasing on
$[0,r_j)$. If $0\les r_j\les x<u$ then
$$
x^{t_{j+1}-t_j}\Cal B^{\,x}_j(g)\ges x^{t_{j+1}-t_j}\Cal
B^u_j(g)>u^{t_{j+1}-t_j}\Cal B^u_j(g),
$$
where the latter inequality obtains since $\Cal B^{\,s}_j(g)<0$
for all $s\in(r_j,\infty)$. Consequently the function $\Cal
B^{\,x}_{j+1}(g)\in\Cal D$ is strictly decreasing. Employing (5.2)
for $x\in[r_j,\infty)$ it follows that $\lim_{x\to\infty}\Cal
B^{\,x}_{j+1}(g)=-\infty$, which completes the inductive step and
the lemma is proved. \qed

\proclaim{Lemma 5.2}Let $g=g_{\Cal C}\in\Cal S\cap\Cal I$ be a
step function investment project, $\Cal C=(C_i)_{0\les i\les n}$.
Let $t_0<s<t$. For all $0<x<y$,
$$
\Cal B^{\,y}_t(g)-\Cal B^{\,x}_t(g)\les c(s,t)\big(\Cal
B^{\,y}_s(g)-\Cal B^{\,x}_s(g)\big)<0. $$
\endproclaim

\flushpar{\bf Proof.} Let $c(u,v)$ be a positive decreasing
accumulation function which is a common lower bound (cf. (5.1))\,:
$0<c(u,v)\les a(u,v),x^{v-u}$; hence also $c(u,v)\les y^{v-u}$. By
introducing cash flows of 0 at times $s,t$ if necessary, one may
assume $s=t_m>t_0$, and $t=t_p$. From Lemma 5.1, $\Cal
B^{\,x}_j(g)\equiv\Cal B^{\,x}_{t_j}(g)$ is a strictly decreasing
function of $x\in[0,\infty)$, $1\les j\les n$. The proof of the
lemma is by induction on $j$ and consists of 3 cases, based on the
iteration scheme (4.2) with respect to the deposit accumulation
function $a(u,v)>0$, the investment accumulation functions
$y^{t-s},x^{t-s}>0$. \itemitem{I:} $\Cal B^{\,y}_j(g)\ges 0$;
$\Cal B^{\,x}_j(g)\ges 0$.
$$
\aligned \Cal B^{\,y}_{j+1}(g)-\Cal B^{\,x}_{j+1}(g)
&=a(t_j,t_{j+1})\big(\Cal B^{\,y}_j(g)-\Cal B^{\,x}_j(g)\big)\\
&\les c(t_j,t_{j+1})\big(\Cal B^{\,y}_j(g)-\Cal
B^{\,x}_j(g)\big)<0.
\endaligned
$$
\itemitem{II:} $\Cal B^{\,y}_j(g)\les 0$; $\Cal B^{\,x}_j(g)\les 0$.
$$
\aligned \Cal B^{\,y}_{j+1}(g)-\Cal B^{\,x}_{j+1}(g)
 &=y^{t_{j+1}-t_j}\Cal B^{\,y}_j(g)-x^{t_{j+1}-t_j}\Cal B^{\,x}_j(g)\\
&\les c(t_j,t_{j+1})\big(\Cal B^{\,y}_j(g)-\Cal
B^{\,x}_j(g)\big)<0.
\endaligned
$$
\itemitem{III:} $\Cal B^{\,y}_j(g)\les 0$; $\Cal B^{\,x}_j(g)\ges
0$.
$$
\aligned
\Cal B^{\,y}_{j+1}(g)-\Cal B^{\,x}_{j+1}(g)&=y^{t_{j+1}-t_j}\Cal B^{\,y}_j(g)-a(t_j,t_{j+1})\Cal B^{\,x}_j(g)\\
&\les c(t_j,t_{j+1})\big(\Cal B^{\,y}_j(g)-\Cal
B^{\,x}_j(g)\big)<0.
\endaligned
$$
\flushpar Note that the case, $\Cal B^{\,y}_j(g)> 0$, $\Cal
B^{\,x}_j(g)< 0$, cannot occur since for all $j\ges 1$ the balances
$\Cal B^{\,x}_j(g)$ are strictly decreasing as a function of $x$.

\flushpar Concatenating the inequalities I, II, III for $m\les j\les
p-1$ ($s=t_m,t=t_p$), one obtains the inequality $\Cal
B^{\,y}_t(g)-\Cal B^{\,x}_t(g)\les c(s,t)\big(\Cal
B^{\,y}_s(g)-\Cal B^{\,x}_s(g)\big)<0$, which completes the proof
of the lemma.\qed

\proclaim{Theorem 5.3} Let $f\in\Cal I$ be an investment project,
minimally supported on $K=[c,d]$. For each $t>c$ the function
$\Cal B^{\,x}_t(f)\in\Cal D$. More precisely, $\Cal B^{\,x}_t(f)$
is a continuous, strictly decreasing function of $x\in[0,\infty)$
such that $\lim_{x\to\infty}\Cal B^{\,x}_t(f)=-\infty$.
\endproclaim

Theorem 5.3 is the main result on general investment projects. It
generalizes the corresponding Lemma 5.1 which treats the
restricted case of step function payment streams that are
investment projects. The proof of Theorem 5.3 involves several
delicate estimates and is given in the Appendix. We now proceed
directly in \S5.2 to the construction of the IRR for general
investment projects.

\flushpar{\bf 5.2. IRR of an Investment Project}

\flushpar Let $f\in\Cal I$ be an investment project, minimally
supported on $K=[c,d]$. Fix a positive (deposit) accumulation
function $a(s,t)$ which is of bounded variation on all compact
intervals $I\subset \bold R$\,: there is a monotone increasing
accumulation function $y(s,t)$ such that $a\les y$. In practice,
$a(s,t)=(1+r)^{t-s}$ where $r$ is an estimated effective interest
rate/year on bank deposits during the life of the investment
project. Employing Theorem 5.3, the balance function $\Cal
B^{\,x}_d(f)\in C^0([0,\infty),\bold R)$, calculated at the end of
the investment project $f$ at time $t=d$, is a continuous,
strictly decreasing function of $x\in[0,\infty)$ such that
$\lim_{x\to\infty}\Cal B^{\,x}_d(f)=-\infty$. In particular the
measure $\nu(f)=m(X_f)<\infty$, where $X_f=\{x\in[0,\infty)\mid
\Cal B^{\,x}_d(f)\ges 0\}$. Thus $\nu(f)=0$ if the function $\Cal
B^{\,x}_d(f)$ is negative, and $\nu(f)=x_0$ if $\Cal
B^{\,x_0}_d(f)=0$ i.e., $x_0\in[0,\infty)$ is the unique root of
the strictly decreasing function $B^{\,x}_d(f)$; if $x<x_0$
($x>x_0$) then the balance $\Cal B^{\,x}_d(f)>0$ ($\Cal
B^{\,x}_d(f)<0)$.

The {\it internal rate of return} (IRR) of the investment project
$f$ is defined to be the effective interest rate/year,
$i_f=\nu(f)-1\in[-1,\infty)$. If $\Cal B^{\,x_0}_d(f)=0$ as above,
then $1+i_f=x_0$ and the corresponding investment accumulation
function at the IRR $i_f$ is $x_0^{t-s}=(1+i_f)^{t-s}$. Note that
$i_f$ depends on the deposit accumulation function $a(s,t)$. The
measure $\nu(f)$ is the relevant parameter, the {\it accumulation
factor}, for computing the IRR/period in the case of investment
projects $f=f_{\Cal C}$, where $\Cal C =(C_j)_{0\les j\les n}$ is
a finite cash flow sequence such that $C_0<0$, cf. Promislow and
Spring (1996). The IRR $i_f$, interpreted as a rate per period
(the time unit is 1 period), coincides with the IRR defined by
Teichroew et al.,\,(1965a) in the special case of discrete
investment projects of constant period whose deposit accumulation
function is $a(s,t)=(1+\alpha)^{t-s}$, where $\alpha>-1$ is a
fixed deposit interest rate per period. In particular, Promislow
and Spring (1996, \S4.2), the rate $i_f$ per period coincides with
the classical IRR in case $f=f_{\Cal C}$ is a loan contract. In
this way, the IRR $i_f$ considerably generalizes the IRR function
defined by Teichroew et al.,\,(1965a) in the case of discrete
investment projects, to the general case of investment projects
defined by payment streams $f\in \Cal I_K$, including the case of
continuous payment streams, under the weak assumption that the
deposit accumulation function is $a(s,t)=e^{g(t)-g(s)}$, where
$g(t)$ can be any function of bounded variation.

\flushpar{\bf Remark 5.6.} Note that the IRR of an investment
project $f\in\Cal I_K$ is robust in the following sense. Let
$f=\lim_{n\to\infty}f_n$, where $(f_n)_{n\ges 1}$ is a Cauchy
sequence of step function payment streams in $\Cal S_K$. Since the
sequence of strictly decreasing balance functions of $x$, $\Cal
B^{\,x}_d(f_n)$, $n\ges 1$, converges uniformly to $\Cal
B^{\,x}_d(f)$ on all compact subsets of $\bold R$, it follows that
$i_f$ is uniformly approximated by $i_{f_n}$ for all $n$
sufficiently large. Furthermore, employing the iteration scheme
(4.2), the balance function of $x$, $\Cal B^{\,x}_d(f_n)$, can be
computed in practice as a finite iteration of T.R.M. balance
functions; hence $i_{f_n}$ is a computable IRR approximation to
$i_f$ for sufficiently large $n$.

\flushpar{\bf Remark 5.7.} Let $f\in\Cal I$ be an investment
project. The rescaled investment project $\lambda f$, for each
$\lambda>0$, has the same IRR: $i_{\lambda f}=i_f$. Indeed, if in
addition $f\in\Cal S\cap\Cal I$ is a step function investment
project then the scale axiom $A_3$ and (5.2) prove that
$i_{\lambda f}=i_f$. The general case follows from Remark 5.6
above. In fact the IRR $i_f$, defined for all investment projects
$f\in\Cal I$, satisfies all of the corresponding postulates for
IRR functions that are presented in Promislow and Spring (1996).

\vskip.5cm \centerline{\bf Appendix}\vskip.5cm

\flushpar{\bf Proof of Theorem 5.3.} Let $f\in\Cal I_K$ be an
regulated investment project minimally supported in $K=[c,d]$, and
let $t>c$. Thus $f=\lim_{n\to\infty}f_n$, where $(f_n)_{n\ges 1}$
is a Cauchy sequence in $\Cal S_K$. Employing (3.3) one may assume
that for all $n$, $f_n\in\Cal S_K\cap \Cal I_K$ is a step function
investment project: the first non-zero cash flow of $f_n$ is $<0$.
Applying Lemma 5.1, for all $n$ sufficiently large (so that the
first non-zero cash flow of $f_n$ occurs in $[c,t)$ and is $<0$),
it follows that the balance $\Cal B^{\,x}_t(f_n)$ is a continuous,
strictly decreasing function of $x\in[0,\infty)$ such that
$\lim_{x\to\infty}\Cal B^{\,x}_t(f_n)=\infty$.

Let $L =[c_1,d_1]\subset[0,\infty)$ be a compact interval,
$d_1\ges 1$. Thus $d_1^{v-u}$ is a positive monotone increasing
accumulation function such that $d_1^{v-u}\ges x^{v-u}$ for all
$x\in L$. Recall that $a(u,v)\le y(u,v)$ where $y$ is a monotone
increasing accumulation function. Applying (4.14) to the common
upper bound $y_1(s,t)=y(s,t)\cdot d_1(s,t)$ $\big(a(u,v)\les
y_1(u,v)$; $x^{v-u}\les y_1(u,v)$ for all $x\in L\big)$, it
follows that for each $t>c$ the sequence of functions, $(\Cal
B^{\,x}_t(f_n))_{n\ges 1}$, $x\in L$, is a Cauchy sequence in the
space of continuous functions $C^0(L,\bold R)$. Consequently the
sequence of functions $(\Cal B^{\,x}_t(f_n))_{n\ges 1}$ converges
to $\Cal B^{\,x}_t(f)$ in the compact-open topology on
$C^0([0,\infty),\bold R)$. It follows that the limit function
$\Cal B^{\,x}_t(f)$ is a continuous, decreasing (i.e.,
non-increasing) function of $x\in[0,\infty)$. However it is not a
formal consequence of convergence in the compact-open topology
that the limit function $\Cal B^{\,x}_t(f)$ either is strictly
decreasing or is unbounded below. We prove below additional
estimates to show that in fact $\Cal B^{\,x}_t(f)$ is a strictly
decreasing function of $x$, assuming $f\in\Cal I$ is an investment
project.

Recall $f\in \Cal I_K$, $K=[c,d]$. Let $0<x<y$, and let $t>c$. We
prove that $\Cal B^{\,y}_t(f)<\Cal B^{\,x}_t(f)$. The proof
divides into two (lengthy) cases. Let $h(u)=y^u-x^u$, $u\ges 0$;
$h(0)=0$. If $y\ges 1$ then $h(u)$ is strictly increasing; if
$y<1$ then $h(u)$ is strictly increasing on $[0,\delta_1]$ where
$h^\prime(\delta_1)=0$ [$\delta_1=\ln\big(\ln x/\ln
y\big)/\ln(y/x)>0$; $\lim_{y\to1^{-}}\,\delta_1=\infty$].

\flushpar {\bf Case I:} $f(c)=0$. Since $f\in\Cal I$ there is a
$\delta>0$, chosen so that also $\delta\les \delta_1$ if $y<1$,
such that $f$ is negative and non-increasing on the interval
$(c,c+\delta]$. One may assume $c+\delta\les t$. Let $\Cal C^n$ be
a cash flow sequence such that $f_n=f_n(\Cal C^n)\in\Cal S_K$. For
all sufficiently large $n$ let the cash flows of the sequence
$\Cal C^n$ in the interval $[c,c+\delta]$ be $C^n_0=0, C^n_1,\dots
,C^n_p$, $p=p(n)$, occurring at times $t_r\equiv
t^n_r\in[c,c+\delta]$, $t_0=c$, $t_p=c+\delta$. Since
$C^n_r=f(t_r)-f(t_{r-1})$, it follows from (3.3) that for all $n$,
$C^n_r\les 0$ for all $r$, $0\les r\les p$. In particular the
partial sums,
$$
S_r=f(t_r)=C^n_0+\dots +C^n_r<0,\quad 1\les r\les p\,; \quad
0=S_0\ges S_1\ges \dots \ges S_p.\tag A.1
$$
Since all the cash flows in the interval $[c,c+\delta]$ are
non-positive, employing the iteration scheme (4.2), it follows
that for all sufficiently large $n$ the balance functions $\Cal
B^{\,u}_{t_r}(f_n)=\sum_{j=0}^rC^n_j\,u^{t_r-t_j}$; $0\les r\les
p$, $u\in[0,\infty)$. In particular at the end point
$t_p=c+\delta\in[c, c+\delta]$,
$$
\Cal B^{\,y}_{t_p}(f_n)-\Cal B^{\,x}_{t_p}(f_n)
=\sum_{j=0}^pC^n_j\,(y^{t_p-t_j}-x^{t_p-t_j}).\tag A.2
$$
We prove that $\Cal B^{\,y}_{t_p}(f_n)-\Cal B^{\,x}_{t_p}(f_n)<0$
and is uniformly bounded away from 0 as $n\to\infty$.

\proclaim{Lemma 5.4} Let $S=a_0u_0+a_1u_1+\dots +a_nu_n$, where
$a_0\ges a_1\ges \dots \ges a_n\ges 0$, and $0\ges S_0\ges S_1\ges \dots
\ges S_n$, where $S_r=u_0+u_1+\dots +u_r$, $0\les r\les n$. Then
$S\les a_mS_m\les 0$ for all $m$, $0\les m\les n$.
\endproclaim
\flushpar{\bf Proof.} By a rearrangement of the terms,
$$
S=S_0(a_0-a_1)+S_1(a_1-a_2)+\dots +S_{n-1}(a_{n-1}-a_n)+a_nS_n.
$$
Since $0\ges S_m\ges S_r$ for all $r\ges m$, and also the successive
differences $a_j-a_{j-1}\ges 0$, it follows that one can replace
each $S_r$ with $S_m$, $r\ges m$, to obtain the inequality,
$$
\aligned
S&\les S_0(a_0-a_1)+\dots +S_{m-1}(a_{m-1}-a_m)+S_m\sum_{j=m}^{n-1}(a_j-a_{j+1})+a_nS_m\\
&=S_0(a_0-a_1)+\dots +S_{m-1}(a_{m-1}-a_m)+S_ma_m\\
&\les S_ma_m.\qed
\endaligned
$$
Since $h(s)=y^s-x^s$, $s\in[0,\delta]$, is  strictly increasing it
follows that the sequence, $(y^{t_p-t_j}-x^{t_p-t_j})_{0\les j<
p}$, is strictly decreasing and positive. Applying Lemma 5.4 to
the sum (A.2), employing also (A.1), it follows that for each $m$,
$1\les m<p$,
$$
\aligned
\Cal B^{\,y}_{t_p}(f_n)-\Cal B^{\,x}_{t_p}(f_n)&\les (y^{t_p-t_m}-x^{t_p-t_m})(C^n_0+\dots +C^n_m)\\
&=(y^{t_p-t_m}-x^{t_p-t_m})f(t_m)<0.\endaligned\tag A.3
$$
One may assume that for some $m \ges 1$, the cash flow $C^n_m$
occurs at time $t_m=c+\delta/2\in(c,c+\delta)$. Employing (A.3) at
time $t_m$, one has the uniform estimate: for all $n$,
$$
\Cal B^{\,y}_{t_p}(f_n)-\Cal B^{\,x}_{t_p}(f_n)
\les(y^{\delta/2}-x^{\delta/2})f(c+\delta/2)<0.\tag A.4
$$
Let the cash flows of the sequence $\Cal C^n$ on the complementary
interval $[c+\delta,t]$ occur at times $t_p=c+\delta,
t_{p+1},\dots ,t_q=t$. Applying Lemma 5.2 ($t=t_q, s=t_p$) and the
inequality (A.4) it follows that for all sufficiently large $n$
(recall that the accumulation function $c(u,v)$ is positive and
monotone decreasing),
$$
\aligned
\Cal B^{\,y}_t(f_n)-\Cal B^{\,x}_{t}(f_n)&\les c(t_p,t)\big(\Cal B^{\,y}_{t_p}(f_n)-\Cal B^{\,x}_{t_p}(f_n)\big)\\
&\les c(L)(y^{\delta/2}-x^{\delta/2})f(c+\delta/2)<0, \quad
L=[c,t].
\endaligned
$$
Passing to the limit as $n\to\infty$,
$$
\Cal B^{\,y}_t(f)-\Cal B^{\,x}_t(f)\les
c(L)(y^{\delta/2}-x^{\delta/2})f(c+\delta/2)<0\tag A.5
$$
Thus (A.5) proves that $\Cal B^{\,y}_t(f)<\Cal B^{\,x}_t(f)$ if
$0<x<y$; hence the decreasing continuous function $\Cal
B^{\,x}_t(f)$ is a strictly decreasing function of
$x\in[0,\infty)$. Furthermore, if $y\ges 1$ then $h(s)=y^s-x^s$,
$s\ges 0$, is an increasing function; hence $\delta$ is fixed,
independent of $y\ges 1$. From (A.5), $\lim_{y\to\infty}\Cal
B^{\,y}_t(f)=-\infty$, which completes the proof of Theorem 5.3 in
Case I.

\flushpar{\bf Case II:} $f(c)<0$. Let $f=\lim_{n\to\infty}f_n$
where $(f_n)_{n\ges 1}$ is a Cauchy sequence in $\SK$, $K=[c,d]$.
Employing (3.3) one may assume that for all $n$, $f_n(c)=f(c)<0$,
hence $f_n\in\SK\cap\Cal I_K$. In what follows we develop an
estimate analogous to (A.5) in order to prove that $\Cal
B^{\,y}_t(f)<\Cal B^{\,x}_t(f)$, where $t>c$ and $0<x<y$.

Let $w=\inf\{1,x\}$ and let $\ep\in\big(0,\frac{w|f(c)|}{2}\big]$.
Since $f$ is continuous on the right, there is a
$\delta\equiv\delta(w)\in(0,1]$, chosen so that also $\delta\les
\delta_1$ if $y<1$, such that for all $u,v\in[c,c+\delta]$: (i)
$f(u)\les f(c)/2<0$; (ii) $|f(u)-f(v)|\les\ep$. One may assume
also $c+\delta\les t$. The construction in \S3.2, property (i),
for the sequence $(f_n)_{n\ges 1}$ of step function approximations
to $f$ shows that for all $n$ one may assume $f_n(t)\in\{f(t_i)\}$
(a finite set), for all $t\in\bold R$, where the cash flows of
$f_n$ occur at times $t_i\equiv t^n_i\in[c,d]$, $c=t_0<t_1<\dots
<t_{q(n)}=d$. In particular for all $n$, \item{(iii)} $f_n(u)\les
f(c)/2<0$ for all $u\in[c,c+\delta]$.

\flushpar Let $f_n= f_n(\Cal C^n)$. For all sufficiently large $n$
let the cash flows of the sequence $\Cal C^n$ in the interval
$[c,c+\delta]$ be $C^n_0=f(c), C^n_1\dots ,C^n_p$, $p=p(n)$,
occurring at times $t_r\equiv t_r^n\in[c,c+\delta]$, $t_0=c$,
$t_p=c+\delta$. Employing (3.3), $C^n_i=f(t_i)-f(t_{i-1})$; hence
applying (ii) ($|f(t_i)-f(t_j)|\les\ep$) one obtains the useful
cash flow estimate in the interval $[c,c+\delta]$:
$$
-\ep\les \sum_{j=k}^rC^n_j=f(t_r)-f(t_{k-1})\les\ep,\quad 1\les
k\les r\les p.\tag A.6
$$
Since $f_n(c)=f(c)<0$ it follows from Lemma 5.1 that for all $n$
the balance functions $\Cal B^{\,u}_j(f_n)\equiv\Cal
B^{\,u}_{t_j}(f_n)$, $j\ges 1$, are strictly decreasing functions
of $u\in[0,\infty)$. With respect to the above data on the
interval $[t_0,t_p]=[c,c+\delta]$, one now proves in addition that
these balances are negative on $[c,c+\delta]$, provided $u\ges w$.
\proclaim{Lemma 5.5} Restricted to the interval $[c,c+\delta]$,
for all $n$, the balances $\Cal B^{\,u}_r(f_n)<0$ for all
$u\in[w,\infty)$, $0\les r\les p$.
\endproclaim

\flushpar{\bf Proof.} Since these balance functions are decreasing
it is sufficient to prove the lemma for $u=w$. Assume inductively
on $r$ that all the balances $\Cal B^{\,w}_j(f_n)<0$, $0\les j\les
r$. The initial balances $\Cal B^{\,w}_0(f_n)=C^n_0<0$, hence the
inductive hypothesis is true at $r=0$. Since all the balances
$\Cal B^{\,w}_j(f_n)$ are negative, $0\les j\les r$, employing the
iteration scheme (4.2), it follows that for all $n$,
$$
\Cal B^{\,w}_{r+1}(f_n)=C^n_0w^{t_{r+1}-t_0}+\dots
+C^n_rw^{t_{r+1}-t_r}+C_{r+1}^n.\tag A.7
$$
Since $w\in(0,1]$ the function $w^t$ is decreasing as function of
$t$. Also for all $r$, the differences $t_r-t_0\les
t_p-t_0=\delta\les 1$. Hence $w^{t_r-t_0}\ges w$, $1\les r\les
p$\,; consequently the first term in the sum (A.7),
$C^n_0w^{t_{r+1}-t_0}\les C^n_0 w$ ($C^n_0=f(c)<0$). Thus
$$
\Cal B^{\,w}_{r+1}(f_n)\les C^n_0w+\sum_{j=1}^{r+1}C^n_j
\,w^{t_{r+1}-t_j}.\tag A.8
$$
Since the sequence $(w^{t_{r+1}-t_j})_{1\les j\les r+1}$ is
strictly increasing (last term is $w^0=1$), Abel's Lemma applies
to the sum (A.8) to obtain, employing also (A.6), the estimate:
for all $n$,
$$
\aligned
\Cal B^{\,w}_{r+1}(f_n)&\les C^n_0w+\sup_{1\les k\les r+1}(C^n_k+C^n_{k+1}+\dots +C^n_{r+1})\\
&\les C^n_0w+\ep\les C^n_0w+\frac{w|C^n_0|}{2}=\frac{C^n_0w}{2}
=\frac{f(c)w}{2}<0,
\endaligned
$$
which completes the inductive step and the lemma is proved. \qed

\flushpar Applying Lemma 5.2 and also the iteration scheme (4.2)
to the balances $\Cal B^{\,u}_r(f_n)<0$ for all $u\ges w$, $0\les
r\les p$, one has the explicit computation for all $n$,
$$
\Cal B^{\,u}_r(f_n)\equiv\Cal B_r(f_n)(a,u)=\sum_{j=0}^r
C^n_j\,u^{t_r-t_j}, \quad 0\les r\les p.\tag A.9
$$
Returning to the proof of Theorem 5.3 in Case II, since $w\les
x<y$ it follows from (A.9) that at time $t_p=c+\delta$, for all
$n$
$$
\Cal B^{\,y}_p(f_n)- \Cal B^{\,x}_p(f_n)=\sum_{j=0}^p
C^n_j\,(y^{t_p-t_j}-x^{t_p-t_j}).\tag A.10
$$
The function $h(s)=y^s-x^s$; $h(0)=0$, $s\in[0,\delta]$ is
strictly increasing. Consequently the sequence
$(y^{t_p-t_j}-x^{t_p-t_j})_{0\les j\les p}$ is non-negative and
strictly decreasing. Applying Abel's Lemma to the sum (A.10), for
all $n$ ($\delta =t_p-t_0$),
$$
\Cal B^{\,y}_p(f_n)- \Cal B^{\,x}_p(f_n)\les
(y^{\delta}-x^{\delta})\sup_{0\les j\les p}(C^n_0+\dots +C^n_j).
$$
Now $f_n(t_j)=C^n_0+\dots +C^n_j$, where $t_j\in[c,c+\delta]$,
$0\les j\les p$. In particular employing property (iii) above, for
all $n$, $f_n(t_j)\les f(c)/2<0$, $0\les j\les p$, and hence one
obtains the uniform estimate: for all $n$ and all $w\les x< y$,
$$
\Cal B^{\,y}_p(f_n)- \Cal B^{\,x}_p(f_n)\les
(y^{\delta}-x^{\delta})f(c)/2<0.\tag A.11
$$
Again as in Case I, let the cash flows of the sequence $\Cal C^n$
on the complementary interval $[c+\delta,t]$ occur at times
$t_p=c+\delta, t_{p+1},\dots ,t_q=t$. Applying Lemma 5.2 ($t=t_q,
s=t_p$) and the inequality (A.11) it follows that for all $n$ the
final balances at time $t$ satisfy the uniform estimate: for all
$n$, if $w\les x<y$ ($c(u,v)$ is positive and monotone decreasing)
$$
\aligned
\Cal B^{\,y}_t(f_n)-\Cal B^{\,x}_{t}(f_n)&\les c(t_p,t)\big(\Cal B^{\,y}_p(f_n)-\Cal B^{\,x}_p(f_n)\big)\\
&\les c(L)(y^{\delta}-x^{\delta})f(c)/2<0, \quad L=[c,t].
\endaligned
$$
Passing to the limit as $n\to\infty$,
$$
\Cal B^{\,y}_t(f)-\Cal B^{\,x}_t(f)\les
c(L)(y^{\delta}-x^{\delta})f(c)/2<0\tag A.12
$$
Thus (A.12) proves that $\Cal B^{\,y}_t(f)<\Cal B^{\,x}_t(f)$ if
$0<x<y$; hence the decreasing continuous function $\Cal
B^{\,x}_t(f)$ is a strictly decreasing function of
$x\in[0,\infty)$. Furthermore, if $y\ges 1$ then $h(s)=y^s-x^s$,
$s\ges 0$, is an increasing function; hence $\delta$ is fixed,
independent of $y\ges 1$. From (A.12), $\lim_{y\to\infty}\Cal
B^{\,y}_t(f)=-\infty$, which completes the proof of Theorem 5.3 in
Case II. The proof of Theorem 5.3 is now complete \qed

\newpage

\centerline{\bf References} \flushpar Arrow, K.J., and D.
Levhari\,(1969): ``Uniqueness of the internal rate of return with
variable \itemitem{}investment.'' {\it The Economic Journal} 79,
560-566.

\flushpar Bourbaki, N. (1949): {\it fonctions d'une variable
r\'eelle.} Paris: Hermann.

\flushpar Dieudonn\'e, J. (1960): {\it Foundations of modern
analysis Vol I.} Academic Press.

\flushpar Donald, D.W.A. (1970): {\it Compound interest and
annuities certain.} Camb.\,Univ.\,Press.

\flushpar Fleming, J.S., and J.F. Wright (1971): ``Uniqueness of
the internal rate of return: \itemitem{}a generalization.'' {\it
The Economic Journal} 81, 256-263.

\flushpar Kellison, S.G. (1991): {\it The theory of interest (2nd
ed).} Irwin, Homewood, Illinois.

\flushpar Norberg, R. (1990): ``Payment measures, interest, and
discounting: an axiomatic approach \itemitem{}with applications to
insurance.'' {\it Scandinavian Actuarial Journal} no.\,1-2, 14-33.

\flushpar Promislow, S.D. (1980): ``A new approach to the theory
of interest.'' {\it Transactions of the \itemitem{}Society of
Actuaries} XXXII, 53-92. \flushpar Promislow, S.D. (1994):
``Axioms for the valuation of payment streams: a topological
\itemitem{}vector space approach.'' {\it Scandinavian Actuarial
Journal} 2, 151-160.

\flushpar Promislow, S.D., and D. Spring (1996): ``Postulates for
the internal rate of return of an \itemitem{}investment project.''
{\it Journal of Mathematical Economics} 26, 325-361.

\flushpar Sen, A. (1975): ``Minimal conditions for monotonicity of
capital value.'' {\it Journal of \itemitem{}Economic Theory} 11,
340-355.

\flushpar Spivak, M. (1980): {\it Calculus (2nd ed.)} Publish or
Perish, Inc. Wilmington, DE (U.S.A).

\flushpar Teichroew, D., R. Robichek, and M. Mantalbano (1965a),
``Mathematical analysis of rates \itemitem{}of return under
certainty.'' {\it Management Science (ser. A)} XI, 395-403.

\flushpar Teichroew, D., R. Robichek, and M. Mantalbano (1965b):
``An analysis of criteria for \itemitem{}investment and financial
decisions under certainty.'' {\it Management Science (ser. A)}
XII, 151-179.

\flushpar Wright, J.F. (1959): ``The marginal efficiency of
capital.'' {\it The Economic Journal} 69, \itemitem{}813-816.
\enddocument